\documentclass[preprints,review,accept,moreauthors,pdftex]{Definitions/mdpi}   

\firstpage{1} 
\pubvolume{1}
\issuenum{1}
\articlenumber{0}
\pubyear{2022}
\copyrightyear{2022}
\datereceived{} 
\dateaccepted{} 
\datepublished{} 
\hreflink{https://doi.org/} 

\newcommand\aj{\textit{AJ}}
\newcommand\apj{\textit{ApJ}}
\newcommand\apjs{\textit{ApJS}}
\newcommand\aap{\textit{A\&A}}
\newcommand\mnras{\textit{MNRAS}}
\newcommand\pasp{\textit{PASP}}

\newcommand{\sevenrm}{\rm\scriptsize}

\newcommand{\OIV}{[O{\sevenrm\,IV}]}
\newcommand{\OIII}{[O{\sevenrm\,III}]}
\newcommand{\OII}{[O{\sevenrm\,II}]}

\Title{Infrared Spectral Energy Distribution and Variability of Active Galactic Nuclei: Clues to the Structure of Circumnuclear Material}

\TitleCitation{Infrared Spectral Energy Distribution and Variability of Active Galactic Nuclei: Clues to the Structure of Circumnuclear Material}

\Author{Jianwei Lyu $^{1}$*\orcidA{}  and George Rieke $^{1}$*\orcidB{}}

\AuthorNames{Jianwei Lyu and George Rieke}

\AuthorCitation{Lyu, J.\& Rieke, G.}

\address{%
$^{1}$ \quad Steward Observatory, University of Arizona, 933 North Cherry Avenue, Tucson, AZ 85721, USA
}

\corres{Correspondence: jianwei@arizona.edu (J.L.); grieke@arizona.edu (G.R.)}

\abstract{
The active galactic nucleus (AGN) phenomena results from a supermassive black
hole accreting its surrounding gaseous and dusty material.  The infrared (IR)
regime provides most of the information to characterize the dusty structures
that bridge from the galaxy to the black hole, providing clues to the black
hole growth and host galaxy evolution.  Over the past several decades, with the
commissioning of various ground,  airborne and space IR observing facilities,
our interpretations of the AGN circumnuclear structures have advanced
significantly through improved understanding of how their dust emission changes
as a function of wavelength and how the heating of the dusty structures
responds to variations of the energy released from the central engine.  In this
review, we summarize the current observational knowledge of the AGN IR
broad-band spectral energy distributions (SEDs) and the IR time variability
behavior covering large ranges of AGN luminosity and redshift, and discuss some
first-order insights into the obscuring structures and host galaxy IR
properties that can be obtained by integrating the relevant observations into a
coherent picture. 
}


\begin{document}

\section{Introduction}

Stars and supermassive black holes (SMBHs) in the form of active galactic nuclei (AGNs) are the 
two main sources of radiation energy in the Universe. Our understanding of stars stands on a foundation 
(the Russell-Vogt Theorem) that their properties to first order depend only on initial composition, mass, 
and age. AGNs, at least with our current understanding, are more complex. Unification theories 
relate the different manifestations of this phenomenon into a general picture. However, the variety of 
observable properties at all spatial scales and wavelength ranges can differ significantly in 
ways not predicted {\it a priori} by theory, nor is their evolution well understood.

A number of reviews of the AGN phenomenon have focused on  
separate aspects. For example, \citet{ulrich1997} focus on the time variability of their outputs; \citet{alexander2012} address the vexing issue of how supermassive 
black holes (SMBHs) grow to their immense masses; \citet{heckman2014} discuss the interactions 
between the evolution of host galaxies and their SMBHs;  \citet{netzer2015} discusses the 
unified model, largely from a theoretical perspective; \citet{padovani2017} present a discussion 
of the unified model across the entire electromagnetic spectrum, with emphasis on the observational 
parameters; \citet{ramos2017} describe the current understanding of how obscuration by circumnuclear 
material alters the observable parameters of AGN; \citet{hickox2018} discuss AGNs that are heavily 
obscured by or even embedded in dense clouds of interstellar gas and dust;  \citet{lacy2020} focus 
on the advances in understanding AGN made with the {\it Spitzer} Space Telescope; and  \citet{cackett2021} 
review the reverberation mapping technique, where time delays in variations of different AGN 
components provide insights to the structure --- they focus largely on the X-ray, UV, optical, and 
broad line regions (BLRs) where this technique has been applied extensively. 

\begin{figure}[htp]
    \begin{center}
  \includegraphics[width=1.0\hsize]{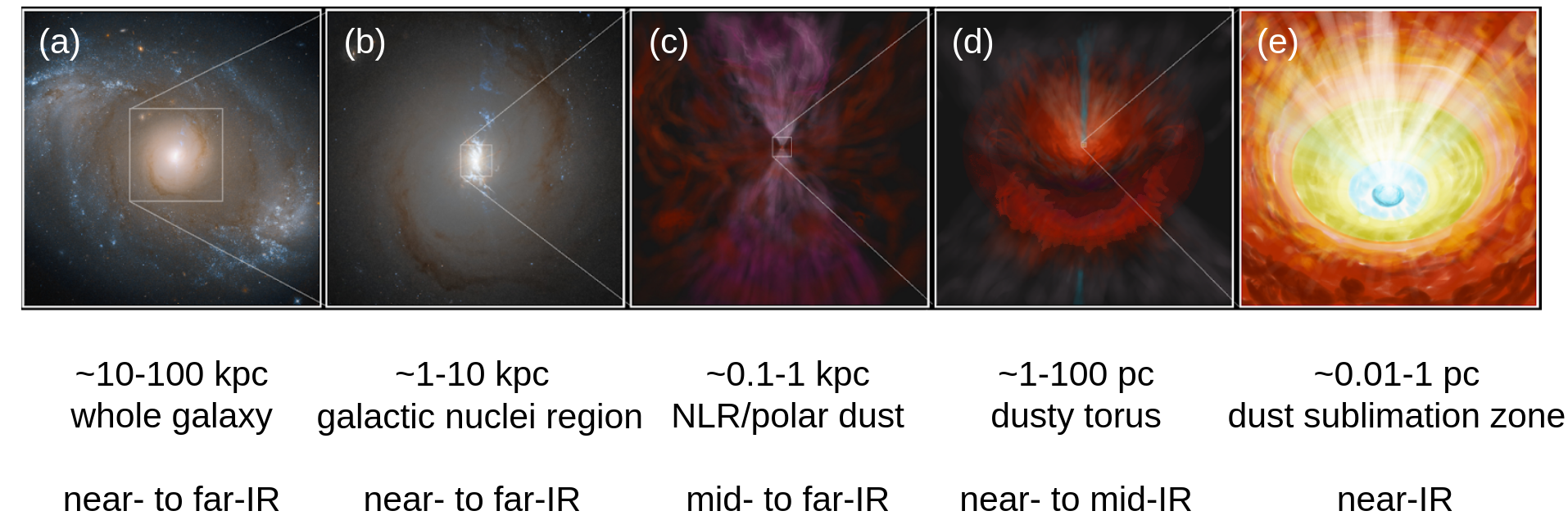}
    \caption{
    Illustration of various components that can contribute to the AGN IR emission at different scales, with the archetypal
    AGN in NGC 4151 as an example. 
    From the left to right, Panel (a): NGC 4151 galaxy image from HST.$^*$ Stars contribute the 
    near-IR ($1 < \lambda <$ 5~$\mu$m) emission and HII regions can  emit strongly in the mid-IR ($5 < \lambda <$ 30~$\mu$m)  and far-IR ($30 < \lambda < 500~\mu$m). Panel (b): the inner 5 kpc 
    region of NGC 4151 as revealed by HST WFC3 images.$^*$ We can directly see the narrow-line-region
    (NLR) clouds near the central bright AGN, projecting along two directions, as well as two dust lanes and 
     numerous small dust streams at different locations. Panel (c): schematic of the dust components at 1 kpc scale, 
     featuring the NLR and polar dust in the AGN ionization cone. Outside this cone, relatively cool dust 
     is shielded by the torus from the direct heating of the AGN. Panel (d): schematic of the dust structures at $\sim$1--100 pc scale, 
     featuring a flared torus with possible turbulent/outflowing signatures and winds feeding the 
     region of the NLR, which often dominate the AGN-heated dust emission from 6--40~$\mu$m. We also indicate a submm torus extension (the circumnuclear disk) on a $\sim$ 40--60 pc scale detected by ALMA around many AGNs \citep{garcia2021}, not observed in NGC 4151 because it is not readily accessible to ALMA. Panel (e): 
     schematic of the innermost region of the torus over $\sim$0.01–-1 pc scales, featuring a likely turbulent 
     disk; here dust grains of  graphite and silicates can be heated to their sublimation temperatures of 1500-2500 K and 900-1000 K respectively, resulting in the AGN hot dust emission peaking around 2--4~$\mu$m. Modified from a figure in \citet{lyu2021}. (* --- figure modified based
     on the original version from \url{https://commons.wikimedia.org/wiki/File:NGC_4151_-_HST.png})
     }
  \label{fig:firstcartoon}
\end{center}
\end{figure}

\begin{figure}[htbp]   
\begin{center}
  \includegraphics[width=0.70\textwidth]{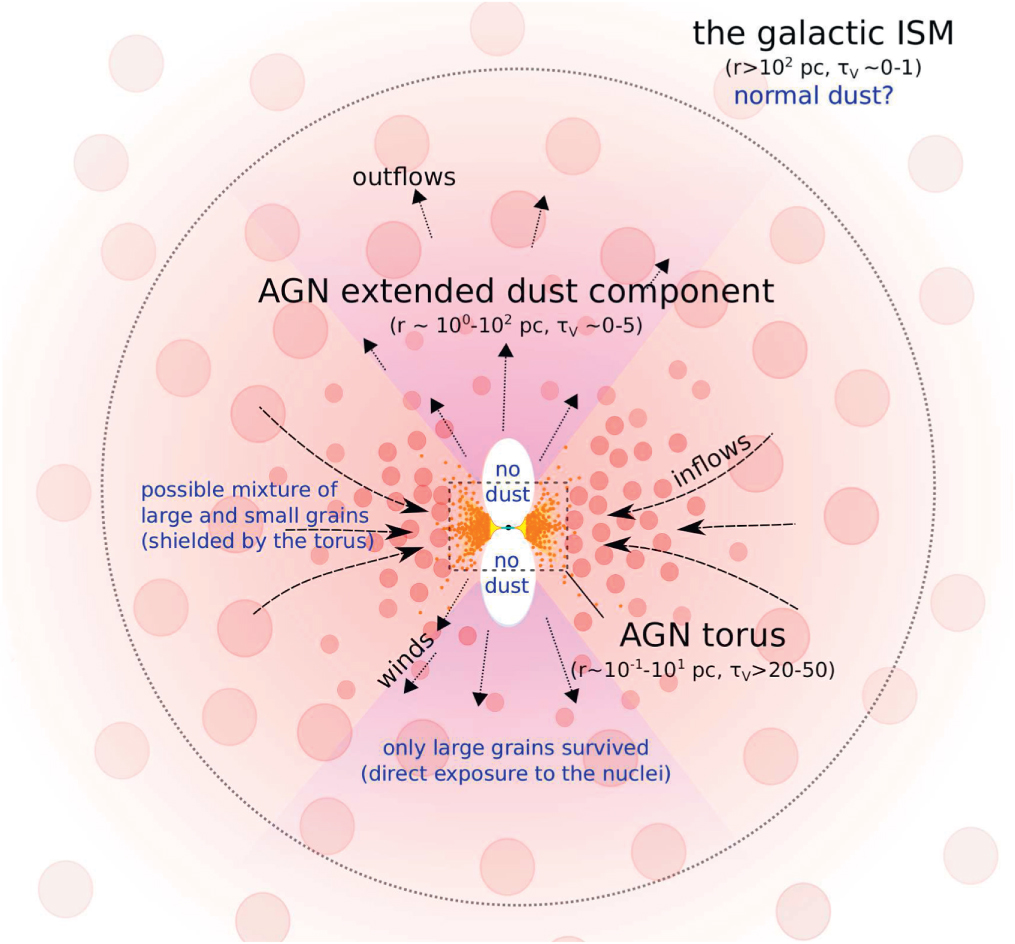}
    \caption{Cartoon illustration of the major IR-radiating components of a classical AGN  \citep{lyu2018}. 
    The central engine and accretion disk are in the center, dwarfed in scale by the IR-emitting 
    components. The circumnuclear ``torus'' is shown in cross section and extends horizontally. We show it 
    as a mixture of a clumpy and continuous distribution of gas and dust, heated by the emission of 
    the accretion disk. The hottest dust, presumably of graphite, is at the inner edge of the torus, 
    within which even carbon dust sublimates.  Perpendicular and on either side of the torus there are 
     biconical outflows where the emission of the accretion disk can escape 
    to excite emission lines. Dust in this region views the accretion disk directly. Close-in it is heated 
    above its sublimation temperature and is destroyed, but further out it can survive and is responsible 
    for ``polar dust emission'' in the mid-IR. In all of these regions, the harsh environment imposed by the 
    intense nuclear ultraviolet radiation \citep{baskin2018}, shocks \citep{dors2021}, and sputtering due to hypersonic drift \citep{tazaki2020}  probably modifies the distribution of dust grain sizes and  
    hence the dust  radiation properties.}
  \label{fig:secondcartoon}
    \end{center}
\end{figure}

Where does this review fit in? The central engines of AGNs are surrounded by a complex set 
of gaseous and dusty structures, which affect the appearance of the AGN in virtually every 
direction. In addition, these structures trace the inflows and outflows 
of material to the central engine, providing important clues to the growth of the central SMBH and
its relation with the host galaxy. 
This review summarizes how our knowledge of these 
structures has grown through extensive infrared (IR) observations and synthesizes the current 
understanding into a coherent picture. As a guide, Figure~\ref{fig:firstcartoon} gives an overview 
of the relevant structures that can contribute to 
the AGN IR emission across different wavelengths, along with their relative scales, as inferred from various current observations. Figure~\ref{fig:secondcartoon} zeros in on the possible 
dusty structures in the immediate region of the central engine of the AGN, whose observed empirical properties 
and their immediate implications will occupy much of our discussion. 

The concepts illustrated in these two figures are largely based on studies 
of the AGN spectral energy distributions (SEDs) and time variability, with additional 
insights provided by IR interferometry and other spatially resolved observations 
of nearby objects. We organize this review around the first two 
approaches, with descriptions when appropriate of how interferometry has upset previous 
concepts and set the thought process on a new path. Section~\ref{sec:agnsed} summarizes
the observed characteristics of AGN IR SEDs, including a brief history, our current ideas of the general AGN IR SED features as well
as possible outliers, and some discussion of IR SED decomposition  and 
the inferred host galaxy properties. Section~\ref{sec:variability} focuses on AGN IR dust 
reverberation mapping and time variability studies, discussing the challenges 
and the results from various campaigns.  Section~\ref{sec:census} stresses the important role of 
AGN IR selection to complete the AGN census with an overview of
selection techniques such as color-color diagrams, SED analysis 
and time variability. We briefly integrate the various AGN IR observations discussed in
this review to provide a synthesis of
the constraints on the AGN dusty environment in Section~\ref{sec:synthesis}.

To clarify our terminology, we use ``AGN'' in a general way to describe any active nucleus whose bolometric 
luminosity $L_{\rm bol}\gtrsim10^8 L_\odot$. ``Quasar''\footnote{
The term ``quasar'' has come into general use beyond the original designation of a radio source and encompasses ``QSO''- quasi-stellar object.} is reserved for the most energetic AGNs with 
$L_{\rm bol}\gtrsim10^{11}~L_\odot$;  the term by itself does not indicate if the nucleus is obscured or not.
Relatively low-luminosity AGNs ($L_{\rm bol}\sim10^8$--$10^{11}~L_\odot$) are termed ``Seyfert nucleus/galaxy''. The word ``torus'' is adopted
to describe the relatively compact optically-thick dusty structures that surround the BH 
accretion disk, although the real geometry can be very different from the classical donut 
cartoon. We describe the more extended structure in the same plane as the circumnuclear disk. ``Polar dust'' refers to the dust distribution in the AGN polar
direction, including dusty winds launched from the torus as well as the more extended dusty NLRs.
We define near-IR as $1 < \lambda < 5~\mu$m, mid-IR as $5 < \lambda < 30~\mu$m, and far-IR as
$30 < \lambda < 500~\mu$m.

\section{The IR Spectral Energy Distribution of AGNs}\label{sec:agnsed}

\subsection{Origin of AGN IR Emission: Thermal or Nonthermal?}
\label{sec:nonthermal?}

The first near- and mid-IR measurements of AGNs, in 1968, showed a roughly power-law spectrum rising 
steeply towards $\sim$ 20 $\mu$m (in the $\lambda- F_\nu$ space) \citep{low1968}. In a prescient paper in 1969, \citet{rees1969} demonstrated 
that this behavior was consistent with heating of circumnuclear dust through absorption 
of emission by a nuclear source luminous in the optical and ultraviolet. Nonetheless, a nonthermal 
explanation for the IR emission thrived alongside the thermal one for about a 
decade \citep[e.g.,][]{burbidge1970, stein1976, neugebauer1979}, bolstered by the 
roughly power-law IR spectral energy distributions. \citet{rees1969} derived 
a minimum timescale for IR variations under their hypothesis that the emission was reradiation of nuclear energy by dust. The first measurements 
of IR variability \citep[e.g.,][]{penston1971} suggested timescales in violation 
of this limit, but these measurements were very early in the development of IR 
astronomy and were subject to significant uncertainties. 

When more complete and detailed IR photometry became available, it was apparent that 
the power-law SEDs were illusory and nonthermal processes no longer needed to be invoked. The observations instead supported thermal models based on the emission from heated 
dust \citep{rieke1978}. In 1981, \citet{rieke1981} reported detailed observations of NGC 4151 that 
confirmed the expected behavior from thermal models, including the role of hot graphite dust  in the near IR and the lack of variability 
at 10 $\mu$m on yearly timescales. In 1987, \citet{barvainis1987} explained the 2 $\mu$m SED bump 
in quasar spectra as arising from the thermal emission of graphite dust heated to its 
sublimation temperature. Two years later, \citet{clavel1989} observed the expected phase 
lag between optical/UV variations and the IR  response to them by heated dust at the expected 
$\sim$ light months distance from the central engine. \citet{hunt1994} reported a search 
for fast timescale variations in the IR similar to those seen in the X-ray and 
concluded that there were none, i.e., the IR variations were consistent with 
reradiation by heated dust. 
The Infrared Astronomical Satellite (IRAS) established that many AGNs emit strongly 
in the far-IR, with SEDs similar to those of non-AGN star-forming galaxies \citep[e.g.,][]{soifer1987}. Given 
this resemblance and the large IRAS beams (1.5$'$ $\times$ 4.7$'$ at 60 $\mu$m, 
~3$'$ $\times$ 5$'$ at 100 $\mu$m), this emission was attributed to the result of 
star formation in the AGN host galaxies.  

These results were persuasive that dust heated by the central engine produces the majority 
of the near- and mid-IR emission for Seyfert galaxies, but that the far-IR probably was powered by young stars. However, the thermal/nonthermal 
ambiguity persisted for higher luminosity AGNs \citep[e.g.,][]{neugebauer1987, kot1992, ulrich1997}. 
\citet{neugebauer1999} conducted  variability monitoring of 25 quasars in the near-IR 
out to 10 $\mu$m; variation at the latter wavelength would be a definitive demonstration that 
the emission was generated nonthermally. However, the signal to noise was inadequate for 
a definitive result. Thus, at the end of the twentieth century, the relative roles of the 
two classes of emission in various high luminosity AGN types --- radio loud/quiet, for example --- were not clear. 
Nonetheless, it was clear that {\it  all} AGNs have a distinctive near- and mid-IR SED compared to stars or galaxies, which we 
discuss below.

\subsection{Behavior of AGN IR Emission. I: General Features}\label{sec:agn-sed-general-feature}

As discussed throughout this review, the form of an AGN SED is a product of how it is viewed and how dust is distributed around it. Type-1 AGNs, by definition viewed pole-on and avoiding heavy dust clouds, are not heavily obscured and hence define the intrinsic SED of this class of source. The general category of obscured AGNs is traditionally termed as Type-2 and defined as not having extremely broad (FWHM$\gtrsim1000$~km/s) permitted emission lines (e.g., H$\alpha$ and H$\beta$). This deficiency results from viewing the source from a perspective such that  optically-thick matter hides the central engine and the BLR. Spectroscopically intermediate cases are designated as Type 1.2--1.9. However, the obscured category is much larger, including many objects so obscured at visible wavelengths that they are too faint to obtain optical spectra. 

\subsubsection{Intrinsic Near- and Mid-IR SEDs}\label{sec:intrinsic}

\begin{figure}[htp]
    \begin{center}
  \includegraphics[width=1.0\hsize]{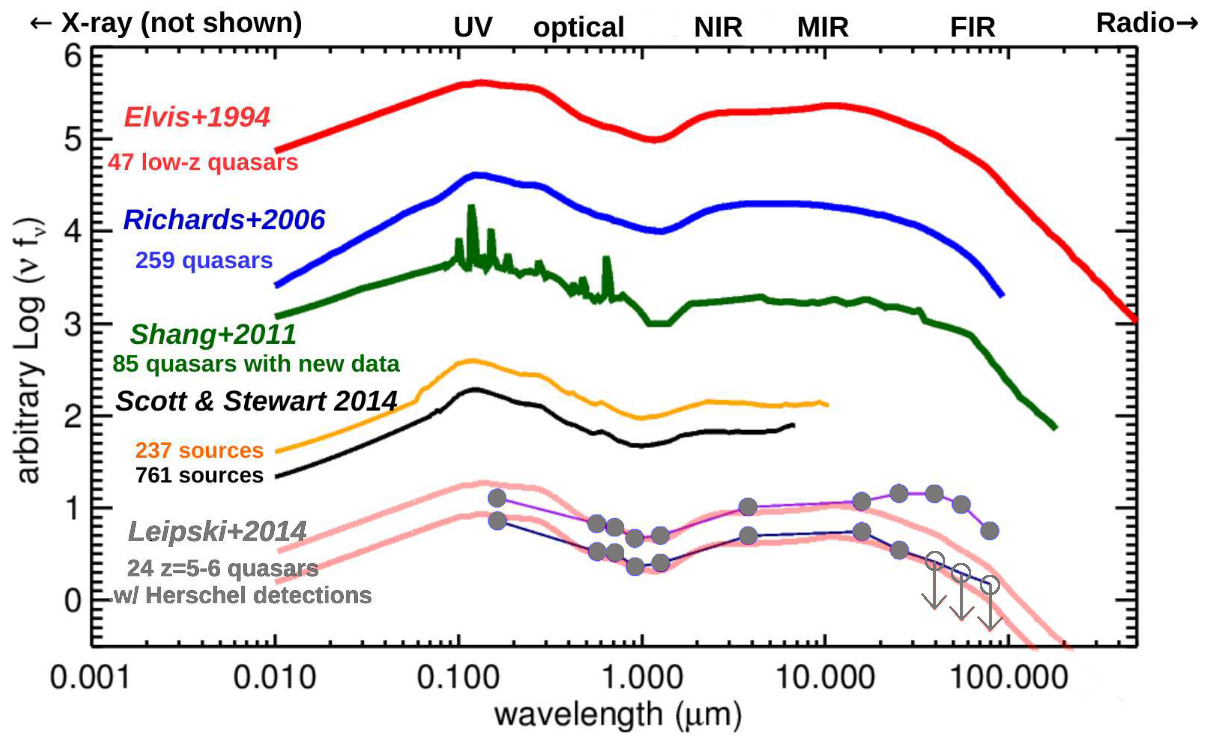}
    \caption{Comparison of empirical average SED templates derived for Type-1 quasars from \citet{elvis1994}, \citet{richards2006}, \citet{shang2011} and \citet{scott2014} as well as the averaged SEDs of $z\sim$5--6.5 quasars with {\it Herschel} detections from \citet{leipski2014}.} 
  \label{fig:quasar_sed}
    \end{center}
\end{figure}

The near-IR SEDs of Type-1 AGNs  were first determined reliably for high 
luminosity AGNs, i.e., quasars, where the AGN emission is so strong it overwhelms 
the stellar contribution in this spectral range \citep{neugebauer1979}.   \citet{ward1987} 
used CCD images to determine and subtract the galaxy contribution to near IR photometry 
of lower luminosity AGNs, i.e., Seyfert galaxies. They showed that the variety of behavior 
was consistent with a single type of SED shape, modified by dust in the form of attenuation 
and IR emission.  A classic paper led by Elvis \citep{elvis1994} proposed a ``standard'' SED 
extending from the IR into the X-ray. As shown in Figure~\ref{fig:quasar_sed}, which also 
shows examples from \citet{richards2006, shang2011, scott2014},  subsequent determinations 
of the SED have been remarkably similar. This similarity extends to very high redshift, as shown 
by the quasar SED at $z\sim$5--6.5 from \citet{leipski2014}. In the near-IR, the compendium of sub-arcsec measurements of Seyfert galaxies 
at $L$ and $M$ bands (3.78 $\mu$m and 4.66 $\mu$m respectively)  by \citet{isbell2021} separates 
the AGN from the galaxy contribution and allows testing this similarity for lower luminosity AGNs. 
Although the wavelength separation of the two bands is too small and the signal to noise too low 
for a determination for each galaxy, a weighted average of the ratio of the $L$ to $M$ band 
fluxes for 22 Type-1 to -1.9 AGNs (cutting two with anomalously large ratios) is 0.9 in 
$\nu f_\nu$, whereas the Elvis template predicts a ratio of 1.01, i.e., the agreement 
is excellent. In addition to these similar slopes, the rapid increase in the AGN near-SED 
near 2~$\mu$m (see Figure~\ref{fig:secondcartoon}) is also seen generally. It is associated with the inner edge of the dusty circumnuclear torus, where 
graphite dust is heated to its sublimation temperature  \citep[e.g.,][]{barvainis1987,mor2012}.  That is, it appears that the shape of the AGN near-IR SEDs illustrated in 
Figure~\ref{fig:quasar_sed} is truly universal and insensitive to AGN luminosity or 
redshift. 
The SED shape at longer wavelengths presumably reflects the structure of this torus, although 
as foreshadowed by the initial attribution of nonthermal emission to apparently power-law SEDs 
(as discussed in Section~\ref{sec:nonthermal?}), the interpretations can be highly degenerate. 

\begin{figure}[htp]
\begin{center}
\includegraphics[width=0.6\hsize]{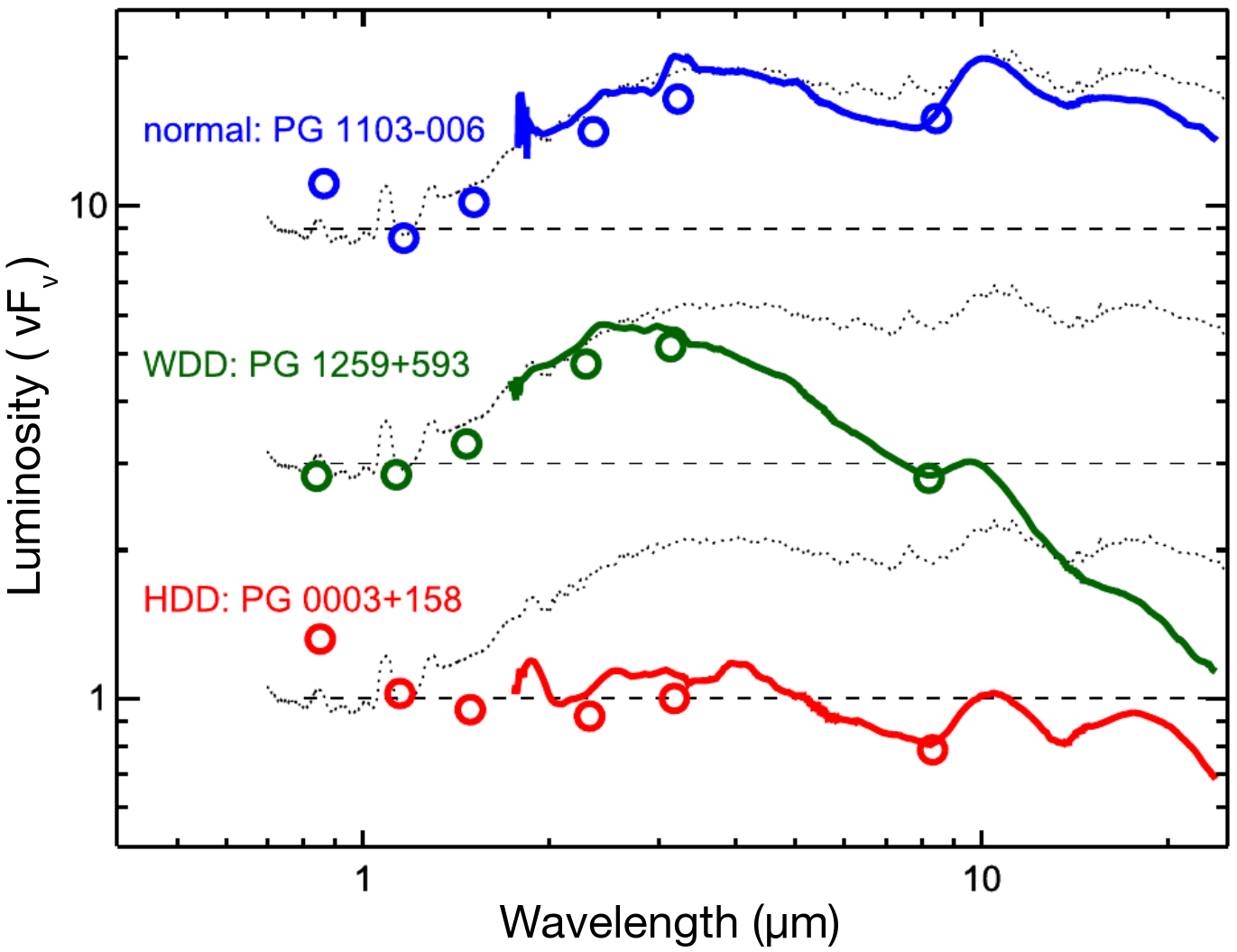}
    \caption{Near- to mid-IR spectra (solid lines) and photometry  (open circles) for three representative Palomar-Green (PG) quasars to illustrate the variations of AGN intrinsic IR emission \citep{lyu2017b}. From 
    top to bottom, PG 1103-006 is a ``normal'' AGN, PG 1259+593 is WDD, and PG 0003+158 is HDD. The dotted lines are the average, i.e. ``normal,'' spectral templates for quasars compiled from the literature,  as in Figure~\ref{fig:quasar_sed}. 
    }
  \label{fig:sed-decomp2}
\end{center}
\end{figure}

Nonetheless, there is some variation in this pattern for a modest number of AGNs.  One notable example is the ``dust-free'' quasars reported firstly at z$\sim$6  \citep{jiang2010}, with SEDs that have a nearly flat continuum from the optical to the mid-IR. However, these objects are not exclusive to very high redshift \citep[e.g.,][]{hao2011}.  A comprehensive study of the AGN intrinsic IR SED variations \citep{lyu2017b} proposes categories of  hot-dust-deficient (HDD: weak emission throughout the near- and mid-IR) and warm-dust-deficient (WDD: weak emission in the mid-IR but similar to those in Figure~\ref{fig:quasar_sed} in the near-IR), which contrast with normal quasars (e.g., those plotted in Figure~\ref{fig:quasar_sed}).  Figure~\ref{fig:sed-decomp2} illustrates the SED characteristics of these types.  Reverberation mapping shows that the near-IR output of even the most extreme HDD quasars is largely due to reradiation of the nuclear emission by hot dust \citep{lyu2019}, i.e., their behavior reflects differences in torus structure, not a change in radiation mechanism. \citet{lyu2017b}  conclude that the HDD quasars constitute $\sim$ 15--20\% of an IR unbiased sample. Later in this review, we will suggest that the ``normal'' category represents an average over all categories, with the HDD and WDD cases as subtypes within this average. There have also been suggestions of systematic intrinsic differences in the circumnuclear tori among Type-1 and Type-2 AGNs (thus their intrinsic SEDs are consequently different), but these claims must be assessed cautiously given the potential for selection biases \citep{elitzur2012}.

\subsubsection{Obscured SEDs} \label{sec:obscured}

Another key aspect in understanding the AGN near- to mid-IR emission is the attenuation by dust. Although some low to moderate extinction occurs for many AGNs, strong attenuation is characteristic of Type-2 or obscured AGNs.\footnote{The classical unified model attributes the obscuration to the circumnuclear torus, but in reality there are many possibilities such as dust in the host galaxy,  dusty clouds not in the torus but along the line of sight, or in extreme cases a dusty cocoon.} In many cases, the
galaxy stellar emission  dominates the near-IR for Type-2 objects \citep[e.g.,][]{hickox2017}, 
making the characterization of the AGN-heated SED behavior at these wavelengths very challenging. 
Two approaches are used. The first, useful at low-$z$,  is to isolate the nuclear emission from
that of the host galaxy with high spatial resolution. For example,
\citet{prieto2010} present arcsec-resolution SEDs of some of the nearest Seyfert-2 galaxies,  Circinus, NGC 1068, NGC 5506, NGC 7582, and build
an average template for Seyfert-2 nuclei.  \citet{alonso2001} and \citet{videla2013} characterize the AGN SEDs at $\sim$1--15 $\mu$m for 
a larger sample (dozens of objects) with the aid of imaging decomposition (as well as some SED decomposition). 
Alternatively, studies have focused on the very bright obscured quasar population
at high-$z$ where the galaxy contamination is relatively weak. This approach includes, e.g.,  the torus template in the SWIRE library \citep{polletta2007}, based on {\it Spitzer} measurements of a heavily obscured but very luminous AGN
at z$\sim$2.5 \citep{polletta2006}, and  average SEDs of IR bright AGNs ($L_{\rm IR}\sim10^{12}$--$10^{13}~L_\odot$) based on a 
sample of $\sim$12 obscured quasars with {\it Spitzer} and {\it Herschel} observations \citep{sajina2012}. 

\begin{figure}[htp]
    \begin{center}
  \includegraphics[width=0.49\hsize]{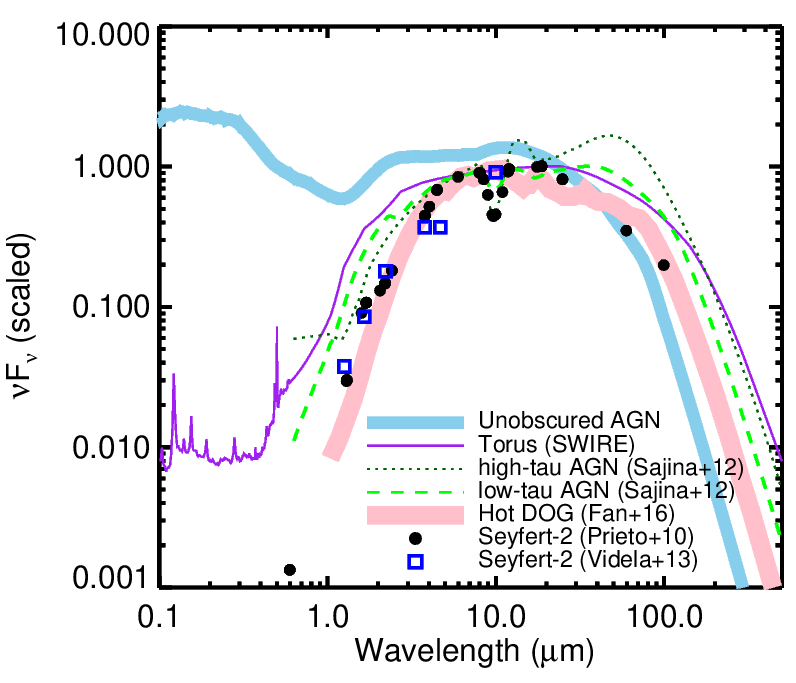}
  \includegraphics[width=0.49\hsize]{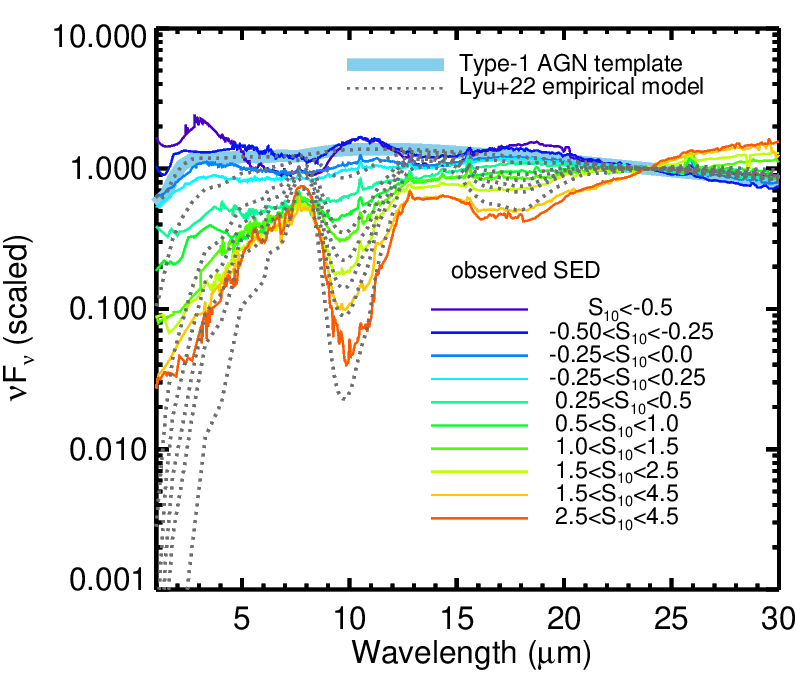}
    \caption{Left: Type-2 AGN SEDs derived in the literature, including
    the average SEDs of Seyfert-2 nuclei derived by \citet{prieto2010} and \citet{videla2013}, the torus template
    in the SWIRE library \citep{polletta2006}, the IR SED templates for high-$\tau_{9.7}$ and low-$\tau_{9.7}$ ($\tau_{9.7}$ respectively $>$ and $<$ 1) IR 
    bright quasars from \citet{sajina2012} and the average SED of hot dust-obscured galaxies from \citet{fan2016}. 
    The SED template of unobscured AGNs from \citet{elvis1994, xu2015a} is also presented as a comparison; 
    Right: The IR SEDs of obscured AGNs at 2--30~$\mu$m. We have computed the average SEDs of AGNs with different 10$\mu$m silicate
    feature strength ($S_{10}=log(F_\textrm{obs.}/F_\textrm{cont.})$=-0.5--4.5) with {\it Spitzer}/IRS spectra and 2MASS photometry (colorful solid lines). The grey dashed lines
    represent the SED model after applying the \citet{lyu2022c}  AGN attenuation curve to the normal AGN template (as defined in  Figures~\ref{fig:quasar_sed} and \ref{fig:sed-decomp2}).}
  \label{fig:observed-sed}
    \end{center}
\end{figure}

In the left panel of Figure~\ref{fig:observed-sed}, we show the obscured AGN IR SEDs from these works \citep{polletta2006, prieto2010, sajina2012, videla2013}, including
the average SED of the so-called hot dust-obscured galaxies, which are believed to host the 
most luminous obscured AGNs \citep{fan2016}. Compared to the unobscured intrinsic AGN template, the SEDs of the  
obscured AGNs have weaker emission with the discrepancy increasing quickly toward shorter wavelengths.
The differences are even noticeable in the mid-IR up to $\sim10$--$20~\mu$m, as evidenced by the silicate absorption features at $\sim$10 and
$\sim$18~$\mu$m.

To reconcile the SED differences of unobscured and obscured AGNs,  efforts have been made to 
generate an AGN attenuation curve.  
Such determinations have centered on the 
optical and ultraviolet (e.g., see a brief summary in \citealt{li2007}), not the near- to mid-IR that is critical to understand Type-2 objects  \citep[e.g.,][]{zafar2015}.
In the IR, the SEDs of obscured AGNs are frequently represented by obscuring a Type-1 AGN template with 
a foreground screen of dust with some standard extinction curve [typically similar to the extinction curve for the Small Magellanic Clouds (SMC)]. However,
the assumption that the extinction curves that work in the UV-optical can be extended into the IR in this simple way is not justified. The AGN accretion disk UV-optical emission and the dust IR emission come from a diverse range of physical scales
and the cause of their obscuration can be quite different (for example, the accretion disk 
can be obscured by the dust along the polar direction while the AGN near-IR emission is obscured
by relatively cold dust in the circumnuclear torus along the equatorial plane). In addition, the properties of dust grains 
surrounding AGNs are likely modified by the harsh environment, as pointed out by, e.g., 
\citet{laor1993} and \citet{maiolino2001}, and the resulting attenuation curve is therefore likely to be unique.   Moreover, the source of obscuration can be also diverse, ranging from the central dusty torus to the host galaxy ISM \citep[see reviews by][]{bianchi2012, hickox2018}.  Given the typically unresolved observations of AGN structure, it is challenging to constrain 
the AGN extinction curve in this spectral range. Nevertheless, \citet{lyu2022c} have used the AGN IR SED variations 
as a function of silicate absorption at $\sim10~\mu$m to generate a near- to mid-IR attenuation curve, based on the behavior from unobscured to heavily obscured AGNs. This attenuation curve can be applied to 
the intrinsic AGN templates (Type-1, Figure~\ref{fig:quasar_sed}) 
to reproduce the SEDs of obscured AGNs reasonably well, as illustrated in the right panel of Figure~\ref{fig:observed-sed}.

\subsubsection{Polar/Extended Dust}\label{sec:polardust}

Spatially resolved imaging reaches resolutions of a few tens of pc for nearby AGNs, using 10-m class telescopes. Emission over 50 pc or more in the mid-IR ($\sim$10~$\mu$m) has been reported for many  
Seyfert nuclei, including Type-2 AGNs in NGC 1068 \citep{braatz1993, cameron1993, bock2000}, Cygnus A \citep{radomski2002}, Circinus \citep{packham2005, reunanen2010}, NGC 1386 \citep{reunanen2010}, IC 5063 and MCG-3-34-64
    \citep{honig2010}, and the Type-1.5 AGN in NGC 4151 \citep{radomski2003}, and there is 
    evidence that it is co-spatial with the AGN NLR clouds traced by the optical 
    \OIII~lines \citep[e.g.,][]{bock2000, radomski2003}.  \citet{isbell2021} and \citet{asmus2016} present
a comprehensive study of sub-arcsec-resolution images of AGNs in the near-IR (3--5~$\mu$m) 
and mid-IR (10~$\mu$m), reporting extended dust structures for many objects. Based on Point-Spread-Function (PSF) 
subtraction on the arcsec-resolution images of 7 AGNs, \citet{fuller2019} found extended 
dust emission over 0.1--1 kpc at $\sim$37~$\mu$m for three AGNs (Mrk 3, NGC 4151, and NGC 4388) that 
might also be attributed to the NLR.

Mid-IR interferometry at 8--13~$\mu$m  with spatial resolution of just a few pc for  nearby AGNs often reveals a polar-elongated dust component, instead of equatorial 
torus-like structures. Together with the more extended polar emission, these components dominate or contribute significantly to the AGN mid-IR emission in both Type-2 
(e.g., NGC 1068, Circinus galaxy, NGC 424; \cite{lopez2016} and references therein)  
and Type-1 objects (NGC 3783 \cite{honig2013}, NGC 5506 \cite{lopez2016}, ESO 323-G77 \cite{leftley2018}). Although these studies are critical for understanding the mid-IR emission, they do not undermine the existence of the circumnuclear torus, since its outer zones may just be shadowed by the inner parts so little energy is deposited in them to be reradiated in the mid-IR.

\begin{figure}[htp]
    \begin{center}
  \includegraphics[width=1.0\hsize]{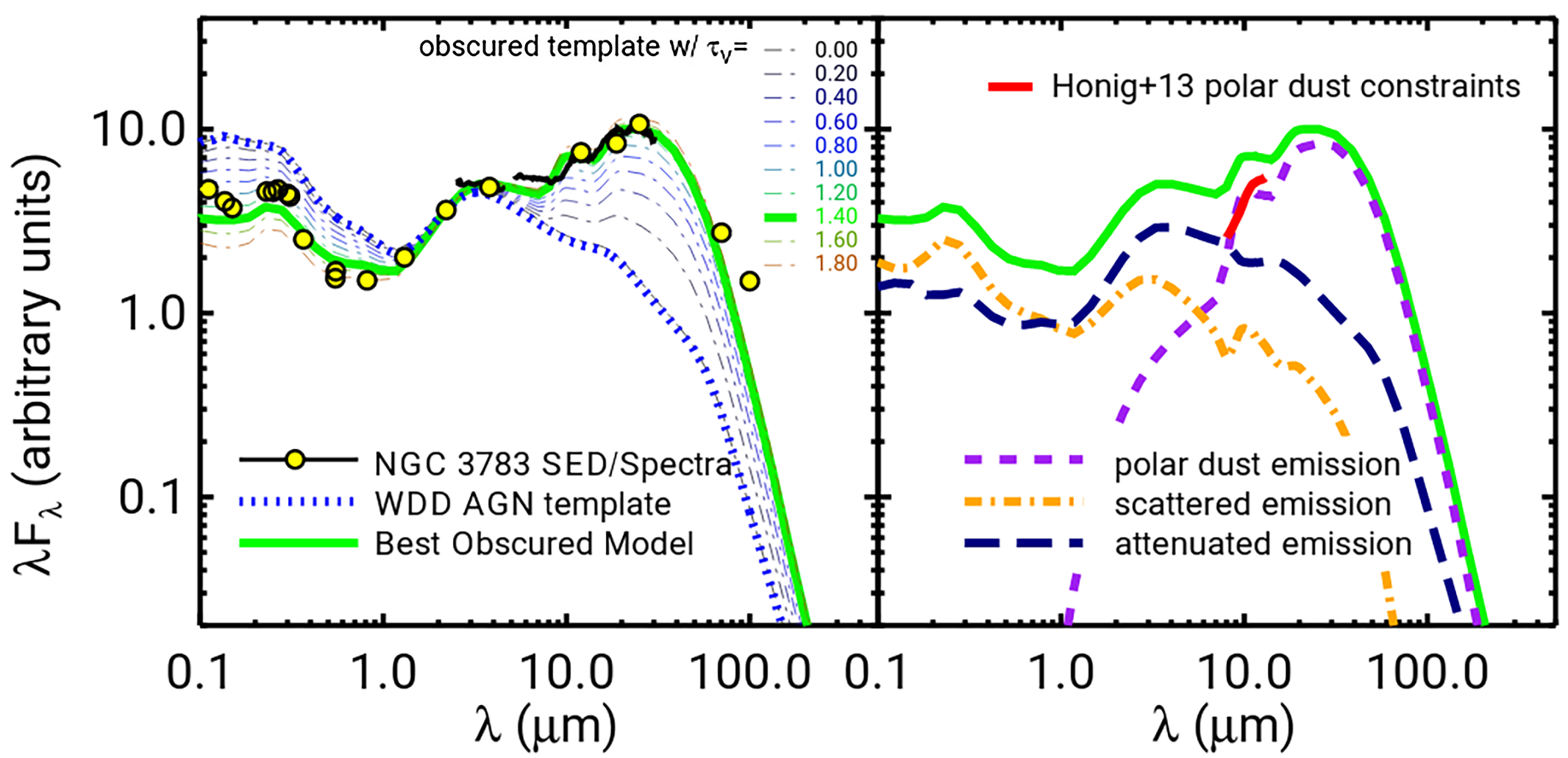}
    \caption{Left: the observed UV-to-far-IR SED of NGC 3783 as traced by high resolution data (yellow circles), with the SED model from \citet{lyu2018}. 
    Right: the contributions of various components of the best-fit model as a function of wavelength. 
    The 8--13 $\mu$m polar dust emission  from  \citet{honig2013} is also shown for comparison. The \citet{lyu2018}
    model obscures the intrinsic AGN template with an extended dust distribution and calculates the resulting SED, which
    includes three components --- (1) attenuated emission of
    the input SED, (2) scattered and (3) polar (i.e. reprocessed) emission from this obscuring dust component, through radiation transfer.  }
  \label{fig:ngc3783}
    \end{center}
\end{figure}

How this polar dust emission behaves as a function of wavelength is poorly known due to the 
lack of multi-band spatially resolved observations. Efforts have been made to 
infer the behavior through SED analysis with some simple assumptions. 
Figure~\ref{fig:ngc3783} shows the SED of 
NGC 3783, which has strong polar emission \citep{honig2013}. Compared with the intrinsic AGN 
templates discussed in Section~\ref{sec:intrinsic}, it has a strong peak in the 20--30~$\mu$m range. 
A similar mid-IR spectral bump is seen in many other Seyfert nuclei \citep{deo2009}. \citet{lyu2018}  attribute 
this feature to  polar dust (e.g., a dusty wind or NLR) and 
propose a semi-empirical model to match the observations. 
They apply the obscuration 
of this extended dust component to the WDD AGN template, showing that this model
 reproduces the integrated SED and the observed mid-IR polar emission strength 
of NGC 3783 (Figure~\ref{fig:ngc3783}). The validity of such a simple model has been further corroborated by fitting 
three more prototypical Type-1 AGNs constrained by interferometry (ESO 323-77, NGC 4507) or 
by high-resolution imaging (NGC 4151), indicating similar properties for these objects. The fitted SED shape among
these sources is very similar \citep{lyu2018}, with a peak near 25--30~$\mu$m.  Most of the SED variations seen among Seyfert-1 nuclei and quasars can be reconciled by adding a similar 
polar dust SED onto the AGN intrinsic templates  \citep{lyu2017b} .

The polar emission is likely to be associated with the NLRs, as suggested by e.g., \citet{netzer1993}. Although there can be exceptions where a strong NLR is not associated with polar IR emission \citep[e.g.,][]{alonso2021}, there are virtually no cases with confirmed polar emission that do not also have strong narrow emission lines. 
Taking advantage of this association, \citet{asmus2019} selected galaxies with strong fluxes in the high-excitation  \OIV  ~25.89 $\mu$m line and found from interferometry that nearly all show polar emission in the 10 $\mu$m range. This led to the suggestion that polar emission is ubiquitous and dominant in the mid-IR for {\it nearly all} AGNs.
However, it is not clear if the \OIV~selection biased the result.  

\begin{figure}[H]
    \begin{center}
  \includegraphics[width=0.6\hsize]{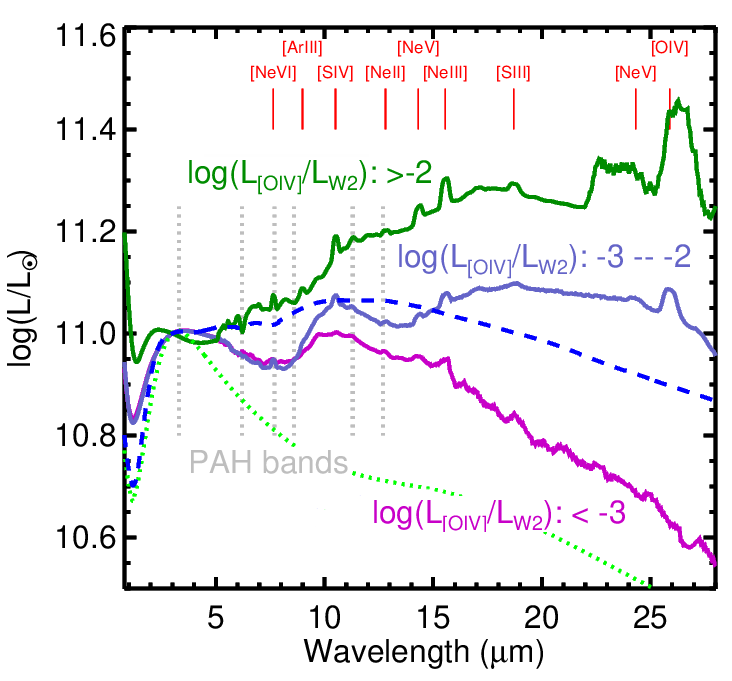}
    \caption{Average AGN mid-IR SEDs for three different bins of the  luminosity in \OIV~relative to that in W2. The green dotted line and blue dashed line represent the WDD
and normal AGN templates, respectively.  All these SED templates are normalized at 3~$\mu$m. Based on \citet{lyu2022b}.}
  \label{fig:oivselection}
    \end{center}
\end{figure}

\citet{lyu2022b}  explore in a more general way if the AGN NLR strength influences the IR SED behavior. 
They trace the relative strength of the NLR by the flux ratio between the mid-IR \OIV~line and the 
WISE W2 ($\sim$4.6 $\mu$m) continuum (the latter traces the hot dust emission luminosity 
at a wavelength where the contamination by the stellar emission is minimized) and studied its relation to 
AGN IR colors such as WISE W2$-$W3 and W2$-$W4. The W3 band ($\sim$ 12~$\mu$m) probes the wavelength range where the existence of polar dust is 
well established and the W4 band ($\sim$22~$\mu$m) is near the suggested peak
of the polar dust SED (Figure~\ref{fig:ngc3783}). They found a strong 
correlation toward larger color differences (more emission in W3 and W4) as the relative \OIV~line strength 
increases, i.e., a correlation between the NLR as traced by \OIV ~and the strength of the emission by polar dust. \citet{lyu2022b} derived average IR SED/spectral templates, as seen in Figure~\ref{fig:oivselection}, which show the dramatic differences in the mid-IR SEDs
as a function of \OIV~relative strength. The suggestion from this behavior is that the ``normal'' 
SED represents an average, including a range from the HDD case to cases with stronger mid-IR than 
even indicated by the normal template. This conclusion is consistent with how the normal template 
is derived --- as an average over a sample of AGNs without selection for individual SED behavior. 

In contrast to spatially-resolved observations of AGN polar dust that are limited to low-$z$ objects 
with preferred inclination angles, 
SED analysis helps us explore the incidence of polar dust among AGNs in general.  
For example, from a study of 64 nearby Seyfert-1 galaxies, \citet{lyu2018} suggested that 
about 1/3 had SEDs consistent with strong emission by polar dust, 1/3 showed 
no evidence for such emission, and the remaining 1/3 were ambiguous. An indication of the role of polar 
dust at higher luminosities can be derived from \citet{xu2015a}, who used the Elvis/normal template to 
fit quasar SEDs and found in some cases that a warm dust component (with strong emission at rest-frame 3--60~$\mu$m) 
had to be added. However, for 37\% of 
their 99 sources, the luminosity in any fitted warm component was $<$ 5\% of the total luminosity 
attributed to the AGN, in agreement, if a bit higher, with the fraction of the Seyfert galaxies with no evidence for 
polar dust.  
The relative strengths of AGN warm dust emission
and NLR emission may both decrease towards higher AGN luminosity \citep[e.g.,][]{maiolino2007, sternJ2012}, indicating
a gradual decrease of the significance of the polar dust component for brighter AGNs. This suggestion is aligned with the increasing
fraction of WDD AGNs \citep{lyu2017b} and the lower  UV-optical extinction seen in quasars compared to Seyfert nuclei 
(e.g. see Figure 6 in \citep{lyu2018}).

\subsubsection{Intrinsic Far-IR Emission}\label{sec:firspec}

\begin{figure}[htp]
    \begin{center}
  \includegraphics[width=1.0\hsize]{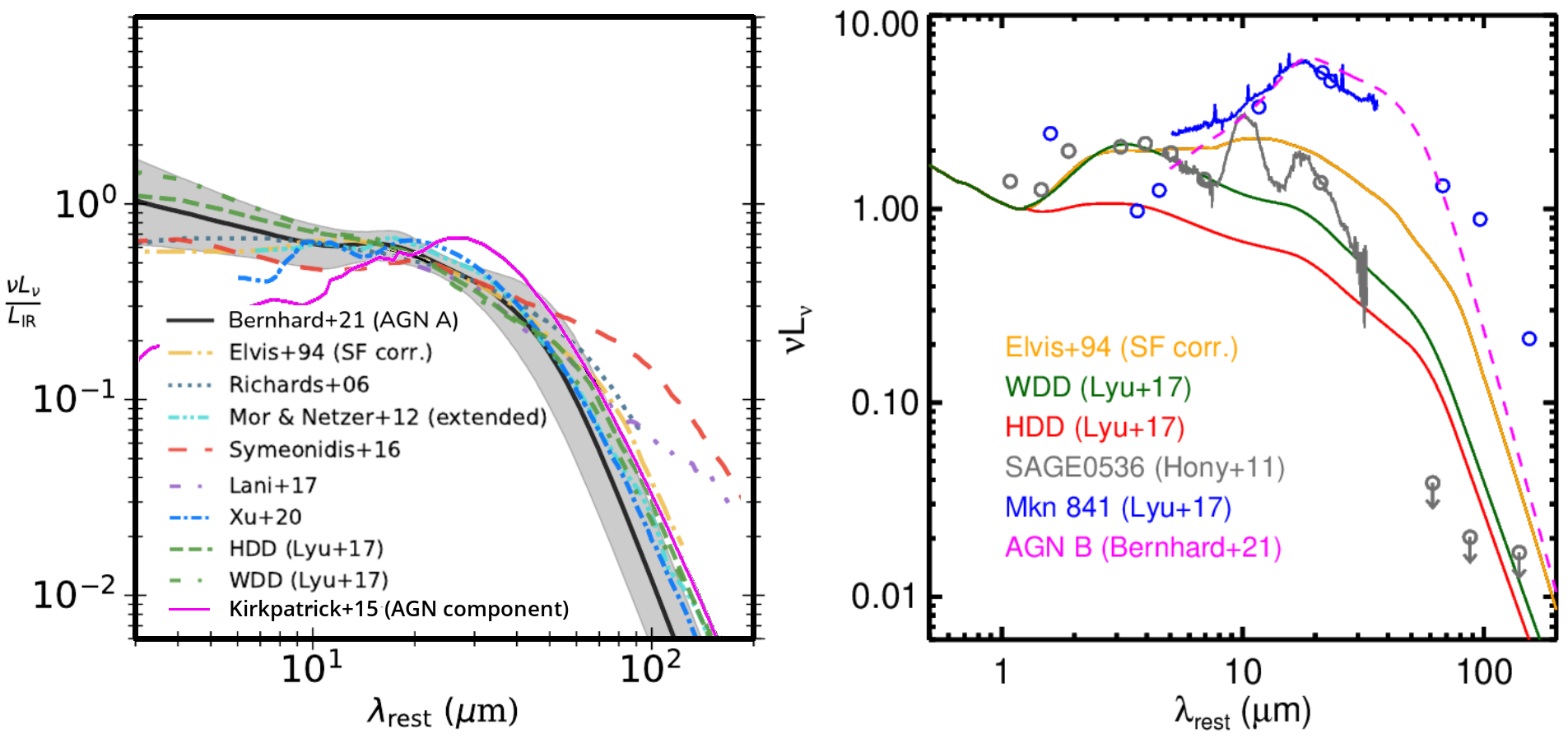}
    \caption{Left panel: comparison of various AGN templates for relatively luminous AGNs or quasars, corrected for host galaxy contribution in the far-IR. These templates are normalized by their IR luminosity. A general agreement is revealed for most papers (modified based on Fig. 14 in \citet{bernhard2021}). Right panel: similar to the left, we only show the normal Elvis-like AGN template \citep{xu2015a}; the WDD and HDD AGN templates \citep{lyu2017a}, normalized in the optical band; the AGN-B template (for less luminous AGNs \citep{bernhard2021}); plus the observed SEDs of AGE1CJ053634.78-722658.5, an extreme AGN at $z\sim0.14$ reported by \citet{hony2011}, and of Mkn 841, a well-studied PG quasar. The spectrum for it includes a substantial contribution from the host galaxy for wavelengths short of 10 $\mu$m; the AGN-only SED is shown by the open circles. 
    }
  \label{fig:fir-sed}
    \end{center}
\end{figure}

The intrinsic far-IR SEDs of AGNs\footnote{By ``intrinsic'' we refer to the product of direct energy input from the AGN, not necessarily to the integrated SED of the entire galaxy.} provide clues to their surroundings on a large scale. However, determining the typical far-IR SED has been challenging. Beyond $\sim$ 30 $\mu$m, there are often substantial contributions from emission by star-formation-heated 
dust in the host galaxies \citep[e.g.,][]{rod1987, wilson1988,rowan1995}, which can contaminate 
the AGN intrinsic SEDs. Removing this contamination is difficult given that the far-IR measurements are made with beam sizes that include both AGN and host. Three approaches have been used to meet this challenge: (1) high spatial resolution on nearby AGNs to isolate the AGN from its host galaxy; (2) use of the aromatic/PAH band(s) to estimate the SFR and its FIR contribution; and (3) use of the FIR SED itself as an estimate of the star formation contribution. Methods of the third type run a risk of being degenerate against the possibility that part of the FIR is powered by the AGN, or that some of the AGN contribution is attributed to star formation.  For example, arguments that galaxies with large ratios of 30 $\mu$m to shorter wavelength mid-IR fluxes are dominated by star formation \citep[e.g.,][]{sajina2012} did not anticipate the ability of AGN polar dust to produce similar behavior.  Although some minor differences do exist, the majority of studies 
 agree that the intrinsic IR SED of most AGNs turns over in the 20--40 $\mu$m range 
and drops quickly into the far-IR \citep[e.g.,][]{miley1985,netzer2007,deo2009, mullaney2011,rosario2012, mor2012, 
xu2015b, kirkpatrick2015, lyu2017a, lani2017, lopez2018, xu2020, bernhard2021} (see left panel in Figure~\ref{fig:fir-sed}).\footnote{Despite this near-consensus, one notable exception
is \citet{symeonidis2016}, who predicted much weaker host galaxy far-IR output from the 11.3 $\mu$m PAH feature 
and thus stronger intrinsic far-IR emission from AGNs. Since its publication, this conclusion has been questioned by a number of independent groups based on studies of the 
same or a similar sample \citep[e.g.,][]{lyu2017a, lani2017, xu2020},
.}

 In a few nearby cases, achieving sufficiently high physical resolution to isolate the AGN emission has been possible  \citep{rieke1975a,rieke1975b,garcia2016, lopez2018,fuller2019}. With the exception of NGC 1068, however, even in these cases an unambiguous separation is not possible past 40~$\mu$m. Frequently in the 30--40 $\mu$m range there is extended emission that can originate either from the NLR/polar dust (and should be included in the FIR SED) or star formation (which should be excluded)  \citep{fuller2019}. The best-determined cases show SEDs that fall off beyond 30 $\mu$m similarly to the quasar SEDs in Figure~\ref{fig:quasar_sed} --- i.e., NGC 1068 \citep{lopez2018,rieke1975a}, NGC 4151 \citep{rieke1975b,fuller2019}. Similar behavior probably also holds for Mrk 3 and NGC 3227, but the combined central and polar SED of NGC 4388 rises at least to 40 $\mu$m \citep{fuller2019}. \citet{garcia2016} used small-aperture photometry on {\it Herschel} data to identify a few additional candidates to have significant AGN-powered output past 40 $\mu$m; we will discuss their results further in Section~\ref{sec:bigfarir}. In general, even with high spatial resolution, it is desirable to have a spectrum to confirm that circumnuclear star formation is not contributing to the flux, since AGNs may be surrounded by compact star-forming regions (see Section~\ref{sec:starformation}).

Many studies of much larger samples have successfully used the 11.3 $\mu$m aromatic band to estimate the SFR and thus the far-IR emission it is responsible for. Subtracting this component from the integrated SED of an AGN plus host galaxy can provide the AGN component. The basis for this approach is that the PAH band provides  an independent measure of the star formation in the host galaxy that allows estimating its contribution to the far-IR output without resolving the nucleus separately.  Typically, a large number of the resulting SF-subtracted SEDs are
averaged to get the final AGN far-IR template(s) \citep[e.g.,][]{netzer2007, xu2015b, symeonidis2016, lani2017, lyu2017a}.

We now elaborate on some cautions for this approach. The carriers of many aromatic features are destroyed around AGNs \citep[e.g.,][]{smith2007}. However, the 11.3 $\mu$m feature is relatively robust in this environment  \citep{diamond2010,esquej2014, alonso2014}. 
There are arguments that it can be excited by the AGN \citep{jensen2017} 
(but at levels generally well below the integrated output of an entire star 
forming galaxy). 
Taking it as a valid measure of the SFR, there are still a number of practical issues in its use for that purpose. 
We illustrate some of them starting with the study of Palomar-Green (PG) quasars by \citet{petric2015} that concluded 
that there was inadequate energy from star formation to power the far-IR from these objects. 

The first issue is simply obtaining an accurate measure of the PAH feature flux. The profiles of the PAH 
features are very broad and blended, leading to two  approaches to measure their strengths:  
(1) a few feature-free continuum points are fitted (typically with linear or spline fits) to determine a local pseudo-continuum above which the core of each individual feature is measured, assuming these features do not have any overlap \citep[e.g.][]{brandl2006}; or (2) a decomposition is employed 
that includes the full PAH features, cores plus wings, as exemplified by ``PAHFIT'' \citep{smith2007}. Typically, the latter approach returns higher values by factors of $\sim$ 2
compared to the measurements based on the simple interpolated continua \citep{smith2007}. Ignoring these systematics 
can undermine the  results.\footnote{To estimate the far-IR SFR from PAH strength,
\citet{petric2015} put PAH strength measurements from \citet{shi2007}, which used the first method, into a formula from \citet{diamond2012}, 
which is derived based on PAHFIT values.  We can estimate the difference by comparing the equivalent widths (EWs) 
for the same objects of the 11.3 $\mu$m feature from \citet{shi2007} with those from \citet{shi2014}, 
since the latter used a procedure similar to PAHFIT. The average ratio of the two EW measurements of the 11.3~$\mu$m PAH 
is 1.74. (also note, \citet{petric2015} incorrectly cite \citet{diamond2010} rather than \citet{diamond2012}.)}
Another concern is the conversion of IR luminosity to SFR; we take the version presented by 
\citet{calzetti2013}, which is a factor of 1.36 lower than that used by \citet{petric2015}. After correcting
the systematics behind the  $L_{\rm PAH}$-$L_{\rm far-IR}$ and $L_{\rm far-IR}$-SFR relations, we find that 
the SFR estimated from the total IR luminosity  is only a factor of 1.26 
greater than that through the 11.3 $\mu$m feature for the sample of \citet{petric2015}. This level of discrepancy could arise because 
the PAH feature lies on the wings of the silicate absorption and hence may be modestly attenuated 
\citep[e.g.,][]{lani2017, lyu2017a, hernan2020}. That is, the estimates agree within errors and 
there is no need for a substantial additional energy source, i.e. the AGN, to power the far-IR in the PG sample. 
This result is confirmed by \citet{xie2021}, 
who use a method based on neon fine structure lines rather than the 11.3 $\mu$m feature to determine 
SFRs for the PG quasar host galaxies and find that the far-IR emission of the PG quasars is generally powered 
by star formation, not by quasar energy.  The above litany of issues based on use of the 11.3 $\mu$m 
feature illustrates the care that must be employed to get accurate SFR values in this application.

The third approach to infer the AGN intrinsic far-IR SED assumes that the observed far-IR emission is largely powered by star formation 
and conducts SED fitting with dust emission templates of purely star forming galaxies 
plus very flexible models of AGN IR emission that are suggested to include realistic AGN SEDs. The AGN model that fits the most observations is then selected 
as the intrinsic AGN template. This approach can be very useful in the mid-IR where the contrast of AGN to 
star-forming contributions is high (as shown, for example, in the identification of AGNs as ``warm'' IRAS sources \footnote{These objects are defined with spectral indices between 25 and 60~$\mu$m of
$-1.25 <\alpha< -0.5$), see the reference for details.}
\citep[e.g.,][]{degrijp1985}). However, it runs the risk of being a circular process for determining the AGN far-IR SED.
Nonetheless, it returns results similar to those using the other two methods  \citep[e.g.,][]{mullaney2011, mor2012, xu2020, bernhard2021}, which supports the conclusion that the far-IR is indeed powered by star formation, as assumed.

ALMA observations in the submillimeter show a circumnuclear disk with an extent
of $\sim$ 40 pc diameter around many AGNs \citep{garcia2021}. From spatially
resolved spectral indices in the submm-mm region, the emission mechanism
appears to be highly inhomogeneous with contributions from thermal emission by
dust, synchrotron emission from jets, and free-free emission
\citep[e.g.,][]{alonso2019,garcia2019}.  It is unlikely, however, that this
component contributes significantly in the mid-IR, since there is no trace of
it in high resolution images such as those discussed at the beginning of
Section~\ref{sec:polardust}. It is dramatically absent in very high resolution
images of NGC 1068  \citep{gamez2022} and the Circinus Galaxy
\citep{isbell2022}.

\subsubsection{Are There AGNs with Substantial Far-IR Output?} \label{sec:bigfarir}
 
 The possibility that some AGNs have substantial far-IR emission on top of the typical SEDs described above 
 has been the subject of a number of  theoretical models \citep[e.g.,][]{ballantyne2006, roebuck2016, mckinney2021}.  Such models are challenged to include realistically the anisotropy of the AGN output and its orientation relative to the host galaxy. In addition, any AGN spectral templates utilized need to be free of contributions from star formation. The latter requirement not only calls for templates that remove star formation effects robustly, as discussed in Section~\ref{sec:firspec}, but can be made more difficult by the increased level of circumnuclear star formation possibly observed in AGNs (see Section~\ref{sec:starformation} below). From an observational perspective, there is not a convincing case that AGNs power a substantial portion of the far-IR in the typical case of host galaxies with high levels of star formation (see Section~\ref{sec:firspec}). 
 Here we discuss whether this situation might nonetheless apply to a subset of AGNs.
 
 Indeed, some cases have been identified where the AGN is claimed to contribute significantly to the flux at 70 $\mu$m \citep{mullaney2011, garcia2016}. For \citep{mullaney2011}, the cause is clearly (their figure 5) that the far-IR contribution from star formation in these galaxies is small and as a result the AGN dominates at 70 $\mu$m by default; the AGN SED turns over as described in Section~\ref{sec:firspec}. For the second reference, we have used the ``normal'' template normalized at 24 $\mu$m to estimate the AGN contribution at 70 $\mu$m. The result is that the AGN may contribute $\sim$ 50\% of the 70$\mu$m ~flux in the central 1 kpc region for three of nine of their candidates for this behavior (NGC 3783, 4151, and 5347), $\sim$ 30\% for IC 5063, and $\sim$ 15\% for three more (NGC 4253, 7213, and 7479). If the AGN contribution at 70 and 100 $\mu$m is removed, the ratio of fluxes at 70 and 100 $\mu$m is slightly bluer in some cases than for main sequence star forming galaxies \citep[based on][]{skibba2011,dale2017}, suggesting that IC 5063, NGC 4253, and NGC 7479 might have a detectable far-IR contribution from the AGN.

 Candidates for this behavior might also appear in  the sample of PG quasars analyzed by \citet{lyu2017a} in the form of a much weaker 11.3 $\mu$m PAH band than predicted by the fits to the far-IR using star forming templates. Among the cases measured with good signal to noise, there are three candidates for this behavior that also have adequate far-IR data to determine SEDs. Two of these (Mkn 841 and PG 1543+489) have spectral drops in the 30--40 $\mu$m range similar to the other AGNs; only PG 1149-110 has a significant far-IR excess above this behavior. 
 
 \citet{netzer2014, netzer2016} have examined this issue for the most luminous known quasars out to z $\sim$ 4.8. They conclude that the far-IR emission in these cases is derived from star formation with, in general, contributions by the AGN being significantly less important. \citet{symeonidis2017} reanalyzed the sample of \citet{netzer2016} and concluded that their far-IR emission was powered by the AGN; however, her conclusion depends on the validity of her quasar SED template as discussed in Section~\ref{sec:firspec}.
 
 Thus, although strong far-IR emission may occasionally occur associated with AGN heating, it must be relatively rare, 
 at least among unobscured sources.  It remains an interesting possibility in heavily obscured cases, as discussed in Section~\ref{sec:iras0857} .  
 An important corollary of the above discussion is that if the far-IR emission 
 of an AGN exceeds the prediction of the normal template significantly, it is very likely to arise from dust heated by 
 recently formed stars (using the far-IR as a SFR indicator should avoid the 5--30 $\mu$m region where the AGN can 
 be dominant).  This means that treating the far-IR of an ensemble of galaxies as being predominantly due to star 
 formation is valid although it may not be for individual  cases, e.g. where the far-IR output is weak.

\subsection{Behavior of AGN IR Emission II: Complications and Possible Outliers} \label{sec:notsimple}

So far, our discussion has focused on ``typical'' AGNs and has shown how their 
SED characteristics can be linked by relatively simple arguments, which is very 
useful in treating a large ensemble of AGNs on the assumption that ``one template applies for all.'' 
There is, however, a menagerie of possible exceptions that we discuss below, indicating that this can be a naive 
premise for some individual objects. 

\subsubsection{Compton-thick AGNs}

Compton-thick AGNs owe their name to having sufficiently large line-of-sight (LOS) gas columns, 
$N_H \ge 1.5 \times 10^{24}$ cm$^{-2}$, to be optically thick to soft X-rays. Applying a 
standard relation between gas column and dust extinction \citep{guver2009}, this corresponds 
to $A_V \ge 7$, i.e., the visible light from the nucleus will be totally lost and the IR SED 
will show very strong obscuration. These sources are nearly invisible from the near IR through 
the soft X-ray, but have received increased attention with NuSTAR hard X-ray observations. Much 
of their energy is absorbed and thermalized and emerges in the mid-IR; a promising spectral 
region to search for them is 4--5 $\mu$m, where stellar photospheric emission from the 
host galaxy is low and PAH bands associated with star formation are absent 
\citep[e.g.,][]{severgnini2012}. The greatly improved IR diagnostics possible with 
the James Webb Space Telescope (JWST) should revolutionize their study. 

\citet{alberts2020} suggest that Compton-thick AGNs can be divided, in roughly equal numbers, into two categories, 
both of which will have similar SED properties. One is cases where the circumnuclear torus is edge-on as we view it, 
blocking the view of the central engine and accretion disk with a very high gas and dust column, but potentially 
identified by NLRs appearing as ionization cones on either side of the central engines. They give 
as an example NGC 5728, studied in detail by \citet{durre2018}. The second is truly deeply buried AGNs whose energy 
is thermalized in the surrounding gas and dust. An possible example is IRAS 08572+3915. This source has extremely deep silicate absorption, 
$\tau_{9.7} = 4.2 \pm 0.1$ \citep{armus2007} but an otherwise featureless continuum, including a complete absence of 
emission lines and aromatic features. The depth of the silicate absorption requires by radiative transfer arguments that the luminosity 
source be buried in the emitting cloud, rather than the emission originating through illumination from the outside 
    the cloud \citep{levenson2007,sirocky2008,nikutta2009}, possibly  indicative of a very deeply embedded AGN (discussion to be continued in the next section).
    
\subsubsection{
Embedded AGNs and FIR Emission}
\label{sec:iras0857}

\begin{figure}[htp]
    \begin{center}
  \includegraphics[width=0.65\hsize]{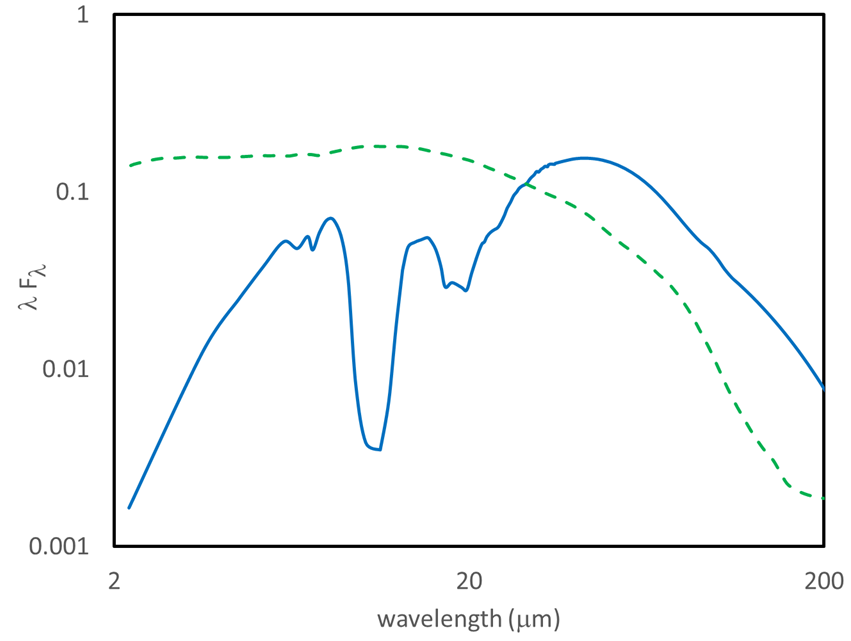}
    \caption{IR SED of IRAS08572+3915 (solid blue line) compared with ``normal'' template (dashed green line). The SED of IRAS08572 has been normalized to half the IR luminosity of the normal template, on the assumption that half the luminosity of the template escapes at shorter wavelengths, whereas all of the shorter wavelength energy is captured and reradiated in IRAS08572, doubling its IR output relative to that of the template. This normalization puts the normal SED at its highest plausible level relative to IRAS08572, but nonetheless there is excess far-IR emission above it.
    }
  \label{fig:iras0857}
    \end{center}
\end{figure}

 As an example of a case where the far-IR may be powered by an embedded AGN, along with the complexities in interpreting the observations, we continue the discussion of  IRAS08572+3915, see Figure~\ref{fig:iras0857}.  Models along the line of the obscured standard AGN SEDs discussed in Section~\ref{sec:obscured} fail to account for the prominent peak in the far-IR, i.e., this object does have substantial far-IR emission above the standard cases, consistent with models such as those of \citet{roebuck2016}. However, \citet{spoon2007,marshall2018} show that a very similar SED can result from  extremely deeply embedded star formation. Nonetheless, \citet{barcos2017} find that IRAS08572+3915 has an extremely compact radio nucleus and substantial IR emission above the predictions of the IR/radio relation. In addition, \citet{iwasawa2011,yamada2021} show it is a hard X-ray source. Thus, it appears that an embedded AGN heats the surrounding ISM to some extent, although a star powered component to the far-IR is still possible. The radio and X-ray measurements are critical to support this interpretation. 

\subsubsection{Hot Dust-Obscured Galaxies}

\citet{eisenhardt2012} discovered a class of extreme sources in WISE photometry, faint or undetected in W1 and W2 
(3.4 and 4.6 $\mu$m) but well detected in W3 and W4 (12 and 22 $\mu$m).  They were confirmed to be AGNs by \citet{wu2012} 
and termed hot Dust Obscured Galaxies (hot DOGs). Subsequent work has found many more such sources \citep[e.g.,][]{tsai2015}. 
A possibly related but less extreme category of extremely red quasars (ERQs) has been reported by \citet{ross2015}.  
Both types of source appear to favor relatively high redshifts, $z > 2$. 
\citet{yan2019} confirm that these sources have huge gas columns ($\ge 10^{25}$ cm$^{-2}$), consistent with their having very hard but  faint X-ray outputs  \citep{vito2018,zappacosta2018}. They are also the sites of large, high-velocity outflows \citep[e.g.,][]{hamann2017, finnerty2020}.

Some modeling of the SEDs of hot DOGs suggests that their emission is dominated by their circumnuclear tori, which 
must have very large covering fractions \citep{fan2016}. Alternatively, it has been suggested that their extreme 
properties result from their very strong outflows or winds \citep{lyu2018, calistro2021}. The relatively large ratio 
of \OIII $\lambda$5007\AA~flux to H$\beta$ \citep{zamamska2016a, yan2019, jun2020} is consistent with the latter possibility.  
It would not be surprising if both hypotheses were relevant. Given their exotic nature, it would be foolhardy to 
be very specific about the circumnuclear environment of these objects until more constraints are available. 

\subsubsection{Suppressed Near IR?}

The Type 1.5 AGN Mkn 841 (=PG 1501+106)  is a  thoroughly studied example with very weak PAH emission. 
Not only is this apparent from the {\it Spitzer} IRS spectrum, but \citet{martin2019} searched for the 11.3~$\mu$m 
PAH feature on sub-arcsec scales and compare with the large-beam measurements with IRS. The feature is not 
detected in either case and they place an upper limit of 0.2 M$_\odot$ yr$^{-1}$ for the SFR in a 400 pc region 
around the nucleus as well as for the entire galaxy. The SED must therefore represent purely the AGN; it is shown in the right panel of 
Figure~\ref{fig:fir-sed}.\footnote{Because the host galaxy has significant photospheric stellar emission (i.e., 
from a relatively old stellar population), we have used the sub-arcsec measurements at $L$ and $M$ from \citet{isbell2021}. 
The $H$-band point is from \citet{peletier1999}; we have subtracted the galaxy as a $n$ = 4 S\'ersic profile 
(sometimes known as a de Vaucouleurs profile); the fit for the inner 3$''$ radius is excellent.}
Compared with normal AGNs, the SED of this object has strong mid-IR excess emission peaked around 20~$\mu$m, 
but the SED still cuts off toward longer wavelengths as for the more typical cases.  
The large rise from the near IR to this peak might be expected for an obscured AGN but is very atypical for a Type-1.5 one.

This behavior might be 
caused by the AGN-heated extended/polar dust, but in that case Mkn 841 is an extreme example. The ratio of fluxes in \OIII $\lambda$5007\AA~to H$\beta$ of 0.62 \citep{winter2010} is relatively high but not extreme, suggesting that the necessary extreme form of polar emission is unlikely. Other geometries are possible, such as a very strong outflow approaching the situation \citet{lyu2018} suggest for hot DOGs. In general, the Mkn 841 type of mid-IR SED is likely to be characteristic of any case where a substantial amount of dust near the nucleus is illuminated directly by the AGN accretion disk.

\subsubsection{AGNs Deficient in 20--30~$\mu$m Output}

In the opposite direction, some AGN SEDs have much less  far-IR emission than  discussed in Sections~\ref{sec:intrinsic}--\ref{sec:firspec}. For example, both HDD and WDD quasars may have considerably 
lower far-IR emission compared to other quasars, as seen in the right panel 
of Figure~\ref{fig:fir-sed}. As noted by \citet{lyu2017b}, normalized by the 
accretion disk luminosity, the HDD AGN has only $\sim$40\% of the dust emission 
at 1.25--1000~$\mu$m of a ``normal'' (or average) AGN and this value for WDD AGN is about 70\%. 
An even more extreme case has been reported by \citet{hony2011} for the galaxy 
SAGE1CJ053634.78-722658.5 at $z\sim0.14$. Although it has similar hot dust 
emission and mid-IR continuum as a WDD AGN, its far-IR emission drops very 
quickly (see the right panel of Figure~\ref{fig:fir-sed}).

\subsubsection{AGN SEDs in the High-Redshift Universe}

To first order, the SEDs of AGNs/quasars at very high redshift ($z\sim 6$) are very similar to 
those at lower redshift, as shown in Figure~\ref{fig:quasar_sed}. For the  examples with {\it Herschel} 
detections from \citet{leipski2014}, the far-IR can very plausibly be associated with star formation in the host galaxies 
\citep{lyu2016}, increasing the case for strong similarities. Similar near- to mid-IR SED variations
such as WDD and HDD are also seen up to these redshifts \citep{lyu2017b}, further supporting the
generally similar behaviors.

It might be expected that very high redshift quasars would lie in low metallicity galaxies and 
possibly be dust-poor, with reduced circumnuclear tori. This was apparently confirmed with the 
discovery of HDD objects in early {\it Spitzer} data \citep{jiang2010}, but the subsequent 
discovery of similar objects at much lower redshift has undermined this explanation. 
Another possibility is that heavily dust-embedded phases in quasar formation and evolution will be 
more common at high redshift, and in this regard the tendency of hot DOGS and extremely red 
quasars (ERQs) to be found at high redshift is intriguing \citep[e.g.,][]{fan2016}, though
efforts have been made to reconcile their SEDs with low-$z$ AGNs \citep[]{lyu2018}. JWST, 
with its enhanced mid-IR photometry capabilities used to conduct a more complete census of 
obscured AGNs, may provide new insights to this possibility. 

\subsubsection{Contamination from Non-thermal Processes}

Besides dust emission, non-thermal processes such as synchrotron emission from a jet 
can also produce near- to far-IR emission, as seen directly in the spatially-resolved 
images of nearby radio-loud systems \citep[e.g.,][]{mcleod1994, uchiyama2006}. According to the classical 
unification model of radio-loud AGNs \citep{urry1995}, the synchrotron emission can be greatly enhanced 
by relativistic beaming if a powerful jet points toward the observer, so this component may dominate
the IR emission in some AGNs.

Some authors have explored the significance of the non-thermal contribution
in the AGN IR emission by spectral/SED decomposition \citep[e.g.,][]{shi2005, cleary2007, westhues2016}. Notably,
\citet{cleary2007} decomposed the mid-IR spectra of radio systems (radio galaxies and radio-bright quasars)
and claimed that the nonthermal contribution can account for 20-90\% of the emission in some systems. However, as pointed out by \citet{haas2008}, such analyses can be problematic given the fitting degeneracies. A more practical solution, as demonstrated in e.g., \citep{shi2005, westhues2016}, 
is to extrapolate the synchrotron emission to the IR, using the  constraints in the radio and millimeter bands that show power laws steepening toward higher frequency. The general conclusion is that the non-thermal emission is not dominant in the IR for most radio-loud AGN, but with some exceptions \citep[e.g.,][]{shi2005}. In fact, the standard SED decomposition
techniques (see Section~\ref{sec:sedfits}) have been successfully applied to study the galaxy and AGN dust-related properties in many radio-bright AGNs \citep{podi2015, westhues2016} and the necessity for a non-thermal component is not obvious for the majority of cases. The generally insignificant contribution of non-thermal emission is also strongly supported by the very similar average IR SEDs of radio-quiet and radio-loud AGNs \citep[e.g.,][]{elvis1994, richards2006, shang2011}. 

Steep spectrum radio sources commonly do not show strong IR variability associated with a 
significant nonthermal source component in the IR \citep{lyu2019}. However, the nonthermal 
IR contribution becomes very prominent for flat spectrum radio sources  \citep[e.g.,][]{lyu2019} 
(see Section~\ref{sec:var-20um}). Since the AGN IR SEDs are typically not observed at the same time, 
the variability of the jet IR emission can cause some non-physical SED features, which complicates 
the interpretations for some radio-bright objects.

\subsection{IR SED Decomposition and Host Galaxy Properties}\label{sec:sedfits}

At some level, the IR SEDs of all AGNs are a combination of contributions by the nuclei 
and the host galaxies. To constrain their properties, SED fitting or decomposition is typically
desired to separate these two components over a wide wavelength range.

\subsubsection{Decomposition of Galaxy SEDs with AGN Contributions}\label{sedfitting}

Normally, there are three dominant components responsible for the integrated IR emission of an AGN and its host galaxy: 
(1) stellar photospheric emission; (2) mid- and far-IR emission by dust heated by young stars; 
and (3) dust heated by the AGN. The first of these components has a nearly uniform spectral 
behavior throughout the IR and can be included robustly in SED fitting 
\citep[e.g.,][]{willner1984, mannucci2001, brown2014}. The second is also 
reasonably uniform in spectral behavior from 1 to $\sim$ 15 $\mu$m where many 
of the challenges in SED fitting are focused \citep[e.g.,][]{dacunha2008,lyuB2022}, 
although there is more variety at longer wavelengths. This behavior is discussed 
further in Section~\ref{sec:SEDinputs}. The mischief lies in the third item.

For the vast majority of AGNs, the existing data provide relatively few truly 
independent constraints for IR SED fitting (where there are multiple measurements, e.g., points in a spectrum, they are often linked by the underlying physics and the number of independent constraints is much less than the number of spectral points).  Thus some AGN SED model or template
is assumed. One popular approach is the adoption of dust radiative transfer models 
under the assumption that the obscuring structures can be described as smooth tori \citep{fritz2006}, clumpy 
tori \citep{nenkova2008} or some hybrid version \citep{stalevski2012, siebenmorgen2015}. 
Such theoretical models have been used by many authors to fit AGN SEDs \citep[e.g.,][]{berta2013, leja2018, shangguan2018, brown2019, yang2020, yang2022, azadi2020, poul2020}.
In general, models of this type have 
more free parameters than the limited number of independent constraints, thus
parameter degeneracy has to be considered (e.g., with the aid of Bayesian fittings). In addition,
these models may lack key features we now understand to be important but where general constraints are still not understood (e.g., emission 
by polar dust). Moreover, our ability to test the validity of 
the theoretical models is limited; in some sense, we have a classic chicken or egg problem because 
the predictions of the models often can only be verified through SED deconvolution. This problem is exacerbated when models representing some kind of average behavior are applied to individual sources without regard to their eccentricities. 
In fact, it has been shown that none of the popular models provide good SED fits to a significant 
subset of AGNs \citep{gonzalez2019}.\footnote{We may find that they have further difficulties 
with obscured AGNs when we have a more complete sample.}

Another useful SED fitting approach is based on 
empirically-determined families of AGN SED templates. Tools have been developed to fit spectra 
\citep[e.g.,][]{nardini2008, hernan2015}, photometry 
\citep[e.g.,][]{calistro2016, lyu2022a}, or a combination of both \citep[e.g.,][]{mullaney2011, bernhard2021}, 
which have also been used over wide ranges of AGN luminosity and redshift 
\citep[e.g.,][]{mullaney2012, xu2015a, lyu2016, lyu2017b, ichikawa2019}.  \citet{ciesla2015} have used a hybrid approach based on theoretical SED models but down-selected to the minimum needed to fit the range of observational parameters. The success of these approaches relies on the challenging process of empirical characterization of the SEDs, including the possible variations, 
of both the pure AGN component and galaxy component.
Although well-conceived approaches of this type must by definition represent the average 
observed behavior reasonably well, they typically do not give insight to the underlying 
causes of that behavior. In addition, if the number of free parameters is restricted, 
the empirical templates can be accurate in an average sense but not cover the 
range of behavior well.

Given the degeneracies and other limitations of SED fitting, it is mainly useful for 
deriving first order properties, such as: (1) AGN SED shape and luminosity (already discussed); (2) host galaxy 
properties directly reflected by the luminosity, such as star formation rates and stellar masses (discussed below); and 
(3) identifying candidates for obscured AGNs (to be discussed in Section~\ref{sec:SEDanalysis}).

\subsubsection{Constraining the Stellar Masses}

Very high resolution imaging can sometimes be used to subtract the AGN emission and measure the flux from the host galaxy, leading to estimates of parameters such as the stellar mass. However, this procedure is often compromised by stray light from the bright AGN component of the image. 

\begin{figure}[htp]
    \begin{center}
  \includegraphics[width=1.0\hsize]{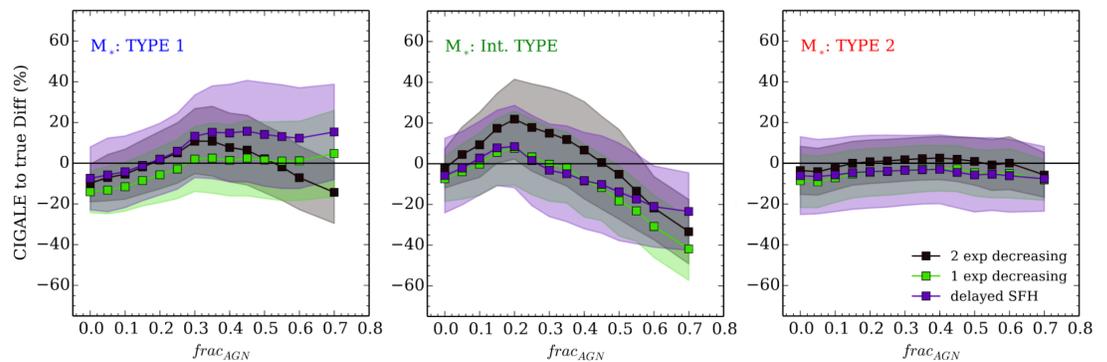}
    \caption{ Fractional differences between the stellar mass derived from SED fitting code {\it CIGALE} and that used as an input to produce the mock photometry from simulations as a function of the AGN-to-total light fraction in the IR luminosity $frac_{\rm AGN}$ (adopted from Fig. 6 in \citep{ciesla2015}, reproduced with permission © ESO.). The galaxy masses are built up through star formation models as keyed to the lower right. See the original paper for details.
    }
  \label{fig:host_mass}
    \end{center}
\end{figure}

An accurate knowledge of the AGN SED allows reasonably good measurements of host galaxy 
properties through SED decomposition. 
In the near-IR, the galaxy stellar continuum is almost identical for different galaxy 
types, with the SED peaking around 1 $\mu$m and dropping quickly following a Rayleigh-Jeans 
tail towards the mid-IR. In contrast, the AGN SED presents a dip around 1.3~$\mu$m with hot 
dust emission peaking at 2--4~$\mu$m. With good wavelength coverage of the integrated 
galaxy emission, we can separate these two components accordingly and retrieve the galaxy 
light fraction at each band. \citet{ciesla2015} have simulated the ability of such SED fitting to return accurate estimates of the stellar contribution of the host galaxies and converted the results to mass using star formation history models. Their study focuses on galaxies with 10 $<$ log(M$_*$) $<$ 11 and considers AGN templates contributing a range of the total IR luminosity from 0 to 70\%. The results are very promising for cases with good photometric coverage, as shown in Figure~\ref{fig:host_mass}. As expected, the derived host galaxy properties are very well constrained for type-2 AGNs, since the optical and near-IR emission is dominated by the stars. However, reasonably accurate values are also returned for type-1 and intermediate cases. For a practical application, \citet{lyu2017b} show that a simple SED decomposition with empirical templates can yield results consistent with dedicated AGN-galaxy image 
decompositions for the PG quasar sample.  With 
an assumed stellar mass-to-light ratio, the AGN host galaxy stellar masses can be reasonably 
determined.  This approach is particularly useful for statistical studies where high-resolution 
image analysis is unavailable \citep[e.g.,][]{xu2015a}.

\subsubsection{Host IR SED Properties}

\begin{figure}[htp]
    \begin{center}
  \includegraphics[width=1.0\hsize]{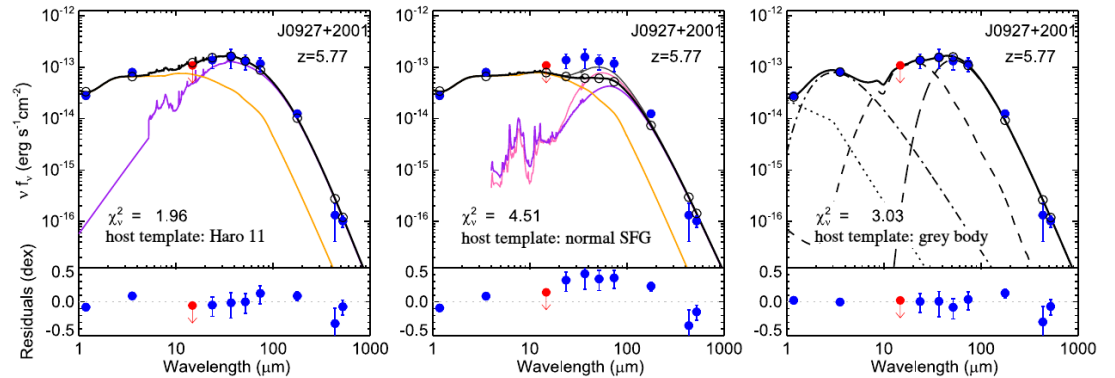}
    \caption{Example SED decomposition of a $z\sim5.8$ quasar. A template based on the local galaxy Haro 11 has been shown to provide a reasonable fit to luminous SFGs at this redshift \citep{derossi2018}.  The left panel shows the decomposition of Haro 11 + normal AGN
    template, the middle panel shows the results of a local SFG + normal AGN template (which works well at low-$z$), and the right
    panel is the result from a complicated SED decomposition model used in \citet{leipski2014}, where the galaxy IR emission is modeled by a grey body with $T=47$~K. The open circles are synthetic photometry performed on the fits, while the blue filled circles are the observed photometry. The Haro 11 fit (left panel) is simpler and equally accurate as the multi-component one, supporting the hypothesis that the far-IR is powered by star formation. This figure is adopted from 
    \citet{lyu2016}.}
  \label{fig:z5quasar-sed}
    \end{center}
\end{figure}

IR SED decomposition is widely used to measure the far-IR luminosity as an estimate of the host 
galaxy star formation, under the assumption that the IR SEDs of these galaxies are the same 
as those of normal star-forming galaxies (SFGs). Empirical 
SFG templates are useful, since more theoretically-based ones contain more free parameters that may result
in over-fitting. Based on AGN and galaxy empirical templates that are well-calibrated against 
statistical observations, standard templates provide a 
satisfactory fit to the far-IR emission of quasar host galaxies at $z\gtrsim0.5$
(\citep[e.g.,][]{netzer2007,xu2015b, lyu2017a,bernhard2021}; see also  Section~\ref{sec:firspec}).  
However, a model with local SFG templates does not work well for quasar host galaxies at $z\gtrsim5$ 
(see Figure~\ref{fig:z5quasar-sed}). Nevertheless, after subtracting the AGN contribution, the far-IR SEDs
of these very high-$z$ quasars resemble those of very luminous purely star forming galaxies 
at similar redshifts \citep{lyu2016,derossi2018}, still consistent with the argument that their 
far-IR emission is largely a result of star formation in the hosts.

\subsubsection{Star Formation} \label{sec:starformation}

SFRs estimated from far-IR luminosities have been used to conclude that quasar host galaxies tend to 
lie on or near to the star-forming galaxy main sequence \citep{santini2012, mullaney2012,rosario2013, berta2013, xu2015b, zhang2016,
dai2018, li2020, hatcher2021}. However, \citet{xie2021} recently concluded that AGN hosts predominantly 
lie above the main sequence. The difference lies almost entirely in the assumed width of the main sequence. 
For example, \citet{xie2021} take it to be $\pm$ 0.2 dex wide, while \citet{xu2015b} take it to be +0.5 dex and $-$1 dex, 
and \citet{zhang2016} take it to be $\pm$ 0.6 dex wide. If we apply the \citet{xu2015b} limits to the $z$=0.2 main 
sequence in \citet{xie2021}, we find that only 14/86 of the  \citet{xie2021} sample lie above it. That is, there is 
general agreement that the quasar host galaxies overall have main-sequence levels of star formation, although 
the sample includes a significant minority of starbursts and  may lie a bit high relative to the central 
main sequence. Even this small bias may be the result of selection effects; the host galaxies of  X-ray selected 
AGNs align more closely with the main sequence with greater scatter around it \citep{ji2022}. Also, there is some 
evidence that AGN hosts lie below the main sequence at low redshifts \citep{barrows2021}.

In Section~\ref{sec:varselect} we mention a sample of AGNs selected via IRAC variability \citep{polimera2018}. 
It was found that the AGN hosts have similar morphologies to those of a control sample, with perhaps a modest 
bias toward disturbed or asymmetric galaxies. A similar result was found by \citet{kocevski2012} using a different 
AGN selection method.\footnote{\citet{kocevski2015} suggest that abnormalities are more frequent for hosts of 
Compton-thick AGNs.} That is, AGN hosts do not seem  preferentially to be  interacting/merging systems \citep{shah2020}.  These results, 
as with the finding that hosts are largely on the star forming main sequence, emphasizes the {\it normality} of 
AGN host galaxies.  

The one possible exception to this rather boring normality is that a number of groups have found evidence that 
the star formation rate in the immediate vicinity of the AGN is often elevated \citep{diamond2012,esquej2014, zhuang2020, dahmer2022}. 

There have been many studies that find that the black hole accretion rate scales with the SFR \citep[e.g.,][]{dai2018}. However, \citet{xu2015b} show that this probably arises because galaxies on the main sequence have a trend of SFR with stellar mass, and the Magorrian relation between the bulge stellar mass and the SMBH mass shows the black hole mass also to grow with stellar mass. Given similar distributions of fractional Eddington luminosities, the range of SMBH masses will be reflected in a similar range of luminosities.  In fact, \citet{xu2015b} show that the slightly off-1:1 correlation of SFR with stellar mass indicated by the slope of the main sequence is reflected in a similar relation for the SFR vs. AGN luminosity. \citet{dai2018} confirm the slope of SFR vs. AGN luminosity that leads to this conclusion.  This is evidence that the apparent linking of the SFR and SMBH  luminosity may arise not through a direct causal connection, but because {\it both} characteristics share a dependence on stellar mass \citep{xu2015b,suh2019}.  

\subsubsection{Do AGN Types Correlate with Different SFRs in Host Galaxies?}

It has been suggested that  the star formation in Type-2 Seyfert host galaxies is systematically higher than in Type-1 hosts. \citet{maiolino1995} found a significant enhancement in the ratio of compact to extended emission for Type-2 hosts by comparing small-aperture groundbased and IRAS measurements at 10.6 and 12 $\mu$m respectively, suggesting that the Type-2 hosts have higher SFRs. Others have found supporting evidence \citep[e.g.,][]{buchanan2006,melendez2008}.   However, \citet{diamond2012} used {\it Spitzer} imaging data and found no statistically significant differences between AGN types in the distributions of nuclear SFRs, extended SFRs, or total SFRs.  Similarly, \citet{masoura2021} studied a X-ray selected sample and also found no significant difference in SFRs with nuclear $N_H$ column\footnote{However, the majority of their SFR estimates were based on WISE photometry, extending only to 22 $\mu$m, which is worrisome because of the potential for AGN contamination}. Observational biases  might explain the apparent differences in results, such as enhanced star formation boosting the brightness of the Type-2 nuclei at 12 $\mu$m and producing an IR sample biased toward this behavior;  anisotropy in the IR emission affecting the 10 $\mu$m outputs of Type 2 AGNs  \citep[e.g.,][]{nikutta2021a}; or  different measures of the SFRs. \citet{zou2019} provide an interesting summary of the contradictions and uncertainties revolving around the issue of systematic type-related differences in AGN host galaxies. 

This issue of differences in SFRs has been examined  with a large sample by \citet{zhuang2020}. They have developed a method using the \OII~and \OIII~optical lines to determine the SFRs. They find a significant tendency toward enhanced SFR in Type-2 hosts. After examining possible selection and other biases, they conclude that their analysis ``suggest[s] that,
on average, the host galaxies of type-2 AGNs have intrinsically higher SFRs than those of Type-1 AGNs.''  \citet{kal2015} used the 250 $\mu$m luminosity as an indicator of the SFR and also found a tendency for type-2 hosts to have higher SFR rates than type-1 hosts, with a formal probability of 0.08 for the two to be drawn from the same distribution for her high signal to noise (at 250 $\mu$m) sample (but a probability of 0.23 for a lower signal to noise sample). \citet{zou2019} have conducted another in-depth examination of this possibility, using far-IR luminosity as an indicator of SFRs. They find a tendency at the 1.4 $\sigma$ level for the hosts of Type-2 AGNs to have higher SFRs than Type-1 ones. They also show that the host galaxies of type-2 nuclei tend to be slightly more massive. These results are all tantalizing but do not establish an effect at very high confidence. The mass difference along with the tendency for main-sequence levels of star formation suggests that any difference might just relate to small systematic differences in host masses.  

To test this latter possibility, we will look for a type-dependent  difference in specific SFR (sSFR), i.e. SFR 
relative to the main sequence. To do so, we use the sample of galaxies with high resolution $L$-band photometry from \citet{isbell2021} (excluding radio bright objects, LINERs, galaxies classified as star-forming, Arp 220, and the Circinus galaxy). Given the close correlation between galaxy stellar mass and the 
luminosity in the 3.6~$\mu$m region \citep[e.g.,][]{wen2013,meidt2014}, we use the integrated 
flux at this wavelength as a proxy for the stellar mass. To derive this flux, we subtract the 
sub-arcsec resolution $L$ and $M$ photometry reported by \citet{isbell2021} from integrated 
$L$-band fluxes for a galaxy (from the NASA Extragalactic Database (NED)), giving the $L$-band 
flux for the bulk of its stellar population.\footnote{We test whether the L-band flux captures 
the integrated output by comparing with the disk flux in the integrated $K_S$ band from 2MASS, 
where we subtract the nuclear emission at $K_S$ according to the subarcsec $L$ brightess and 
the normal template for Type-1 and 40\% of this value for Type-2 (the selection of values 
between 0 and 100\% had only a $\sim$ $\pm$ 5\% effect on the average and median values). 
We correct to the expected $K_S$ to $L$ flux ratio if there is a deficiency.}  Next, the 
luminosity at 60--100 $\mu$m correlates closely to the total IR luminosity  \citep[e.g.,][]{rieke2009} 
and we can use it as a proxy for the star formation rate. Thus, the ratio of the 70 and 3.5 $\mu$m 
luminosities, or for galaxies at low redshift, of the fluxes, becomes a proxy for sSFR.  
In this sample, the distribution of this ratio is offset by 0.3 dex higher for Type-2 hosts than for Type-1 ones; the Kolmogorov-Smirnov probability that these two 
distributions are drawn from the same population is 12\%.   This result and those by \citet{zou2019,zhuang2020}  are of modest statistical significance, but their suggestion of an effect is worthy of further investigation.

\section{Probing AGN Dusty Structure with IR Reverberation Mapping and Variability}\label{sec:variability}

 Initially it was hypothesized that AGN IR emission might be generated nonthermally (Section~\ref{sec:nonthermal?}). 
The first observations of IR variability in AGNs addressed this issue, testing for variations sufficiently rapid to require a nonthermal origin. The demonstration of  phase lags in the IR relative to variations in the blue and UV, as predicted for a thermal origin, largely settled this issue \citep[e.g.,][]{clavel1989,sitko1993}.  Attention then turned to using the observed time delays between nuclear variations and changes in the IR as circumnuclear dust absorbed and reradiated the nuclear emission  --- ``reverberation mapping'' --- to learn about the structures of the  material surrounding the AGN central engines. 

\subsection{Challenges}

 Reverberation mapping is applied extensively in the
X-ray, ultraviolet, optical, and for both continuum and emission lines, where it has revealed aspects of the structure of the regions emitting in these spectral ranges (see a recent review by \citet{cackett2021}). From this experience, achieving unambiguous results requires a measurement sequence that samples the variations at the Nyquist rate, with this sampling extending over as a minimum three times the longest reverberation
lag, although modest gaps in longer datasets should not strongly affect the results \citep[e.g.,][]{horne2004}. With limited datasets, even longer monitoring is needed, e.g. 10 times the longest period \citep{kozlowski2017}. This degree of sampling is relatively easy to achieve in the ultraviolet or optical, where the relevant lag  timescales are days or at most weeks. However, the inner edges of the circumnuclear tori in Seyfert galaxies are sufficiently distant from the central engines that minimum lags of months are typical, and the lags for quasar luminosities can be years. This lag length makes it challenging in conventional ground-based  campaigns to observe at high cadence over three or more cycles. Shorter sequences, or less intensive sampling, can be useful only if the free parameters in the lag determination are severely limited. 

\begin{figure}[htp]
    \begin{center}
  \includegraphics[width=1.0\hsize]{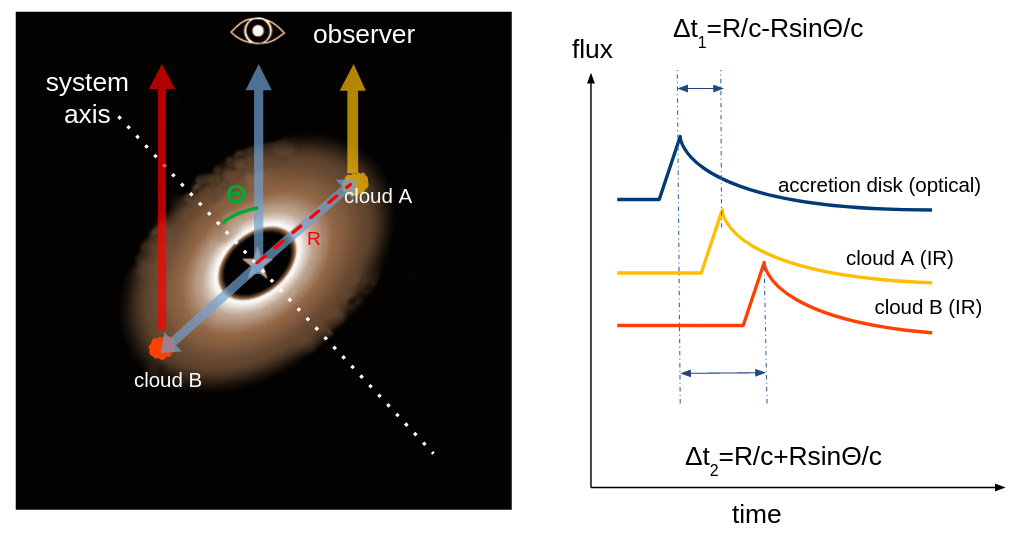}
    \caption{Illustration of the basic concepts of AGN dust reverberation mapping. Assuming two 
    identical dusty clouds (the red and yellow clouds in the figure) at the same distance, $R$, to the 
    central engine (the blue star), their IR variability signals will be relatively delayed to the optical flare 
    from the accretion disk by $\Delta t_1=(R/c)(1-\sin\theta)$  (yellow lines) and $\Delta t_2=(R/c)(1+\sin\theta)$ (red lines), 
    where $c$ is the speed of light and $\theta$ is the viewing angle of the observer relative to the 
    system axis. In a more realistic case, the IR reverberation signals seen in one band come from a composite of clouds
    at some ranges of distance and scale height with some possible radiation transfer effects.
    }
  \label{fig:laggeometry}
    \end{center}
\end{figure}

A second issue is general to reverberation mapping. The extent of the emitting region along the line of sight smooths the lag and blurs features that might be indicative of structures, as illustrated in Figure~\ref{fig:laggeometry}. An example might be a circumnuclear torus that lies with its plane {\it along} the line of sight. The part of the torus closest to the observer would then show virtually no lag, while that furthest from the observer would show a lag indicative of twice the radius of the torus. Other parts of the torus would fill in between these extremes. That is, the lag itself would be appropriate for the size of the torus but the smoothing could make it impossible to deduce much more about the structure. One way to mitigate this issue is to focus on Type-1 AGN, where one expects the circumnuclear torus to lie closer to the plane of the sky than in Type-2 objects. 

\subsection{Lag Measurement Campaigns and Basic Results}\label{sec:rm-results}

A pioneering and very extensive groundbased near IR AGN monitoring effort was conducted at the Crimean Observatory of the Sternberg Astronomical Institute (SAI) from  1994 through 2018 \citep{lyutyi1998,shenavrin2011, taranova2013, oknyansky2018}, with some monitoring extending back to 1985 \citep{lyutyi1998}. These measurements used an aperture photometer with an InSb detector, allowing not only the $J$ (1.25 $\mu$m), $H$ (1.65 $\mu$m), and $K$ (2.2 $\mu$m) bands to be covered, but also the $L$ band (3.6 $\mu$m). To mitigate the influence of atmospheric seeing on aperture observations  \citep{peterson1995}, the measurements were made with a large aperture (12$''$), requiring an accurate correction for the galaxy emission within this aperture. Because of the long extent of the measurement series, including phases where the AGN was very faint, accurate corrections were possible \citep[e.g.,][]{taranova2013}.

The Multicolor Active Galactic Nuclei
Monitoring (MAGNUM) project \citep{yoshii2003} used a photometer specifically designed for reverberation-type observations. It  fed the visible light to a CCD and the IR to an InSb detector array, allowing simultaneous observations from the $U$-band (0.36 $\mu$m) through the $L'$-band (3.6 $\mu$m). It was used to monitor a number of Type-1 Seyfert galaxies and quasars in the $V$-,$R$-, $I$-, $J$-, $H$-, and $K$- bands  \citep[e.g.,][]{minezaki2004, suganuma2006} from 2001 through 2007, with the results summarized in e.g., \citet{koshida2014} and \citet{minezaki2019}. In principle, the use of an IR array should improve the robustness against seeing fluctuations and allow subtraction of the galaxy contribution by profile fitting (but for the latter, \citet{lyu2021} find that the galaxy is under-subtracted, possibly due to a strong central concentration of the stellar population). 

In addition to these two large-scale reverberation mapping campaigns, there are a large number of focused studies of individual, or a few, sources utilizing near-IR photometry out to the $K$-band, including e.g., 3C273 
\citep{sobrino2020}, NGC 5548 \citep{landt2019}, MCG 6-30-15, \citep{lira2015}, NGC 3783 and MR 2251-178 \citep{lira2011}, PGC 50427, \citep{nunez2015}, 3C120 \citep{ramolla2018}, H0507+164 \citep{mandal2018}, and Z229-15 \citep{mandal2021}. $K$-band measurements are well-suited to determining the behavior of the hottest dust ($\sim$ 1500 K) at low redshift, but to extend the study to higher redshifts and/or lower-temperature dust, it is essential to monitor at longer wavelengths. Of the ground-based observations discussed so far, only the Sternberg Astronomical Institute campaign goes to $L$-band at 3.6 $\mu$m. With  {\it Spitzer}/IRAC observations of Seyfert-1 nucleus in NGC 6418, \citet{vazquez2015} report dust time lags at 3.6 and 4.5~$\mu$m and find they are roughly consistent with the K-band reverberation results. 

The WISE/NEOWISE mission, which completes an all-sky survey roughly every 6 months in two photometric bands at 3.4 and 4.6 $\mu$m, offers the opportunity to
conduct statistical AGN near-IR reverberation mapping with the combination of optical time-domain surveys on the ground, as firstly demonstrated by \citet{lyu2019}. The low cadence of the IR light curves requires that reverberation mapping using these data must be focused on high luminosity objects, where it is expected that the circumnuclear torus is large and hence the reverberation timescales are long. Some representative studies using these data include: e.g., PG 1302-102 \citep{jun2015}; 87 PG quasars \citep{lyu2019}; $\sim$ 690 quasars in SDSS Stripe 82 \citep{yang2020}. The latter two studies use different approaches to smooth and compare the visible and IR light curves, but they obtain very consistent results \citep{yang2020}. \citet{lyuB2022} studied the IR reverberation properties of 13 changing-look AGNs and found they follow a  similar relation defined by the PG quasars \citep{lyu2019}. Finally, \citet{noda2020} use X-ray and NEOWISE data on the Type-2 AGN NGC 2110, finding a lag consistent with those for Type-1 objects.

\begin{figure}[htp]
    \begin{center}
  \includegraphics[width=0.8\hsize]{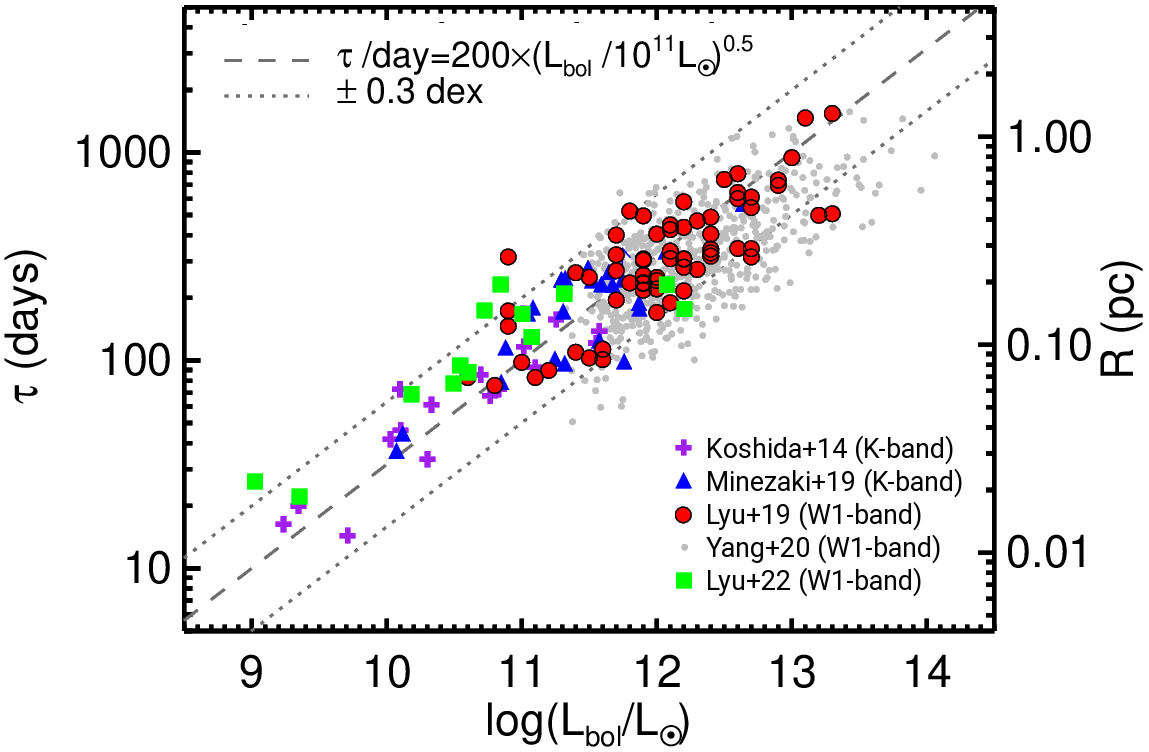}
  \caption{
Correlation between the near-IR dust time lags and AGN bolometric luminosity. These measurements include the ground-based K-band ($\sim2.2~\mu$m) reverberation mapping results of $\sim$ 50 Seyfert-1 nuclei and quasars from the MAGNUM project   \citep{koshida2014,minezaki2019}, and the WISE W1-band ($\sim3.4~\mu$m), results for 87 PG quasars \citep{lyu2019}, $\sim590$ SDSS Stripe-82 quasars \citep{yang2020}, and 13 changing-look AGNs \citep{lyuB2022}. Despite the various uncertainties and different systemics in these works, there is a strong correlation between the dust time lag and AGN luminosity over five orders of magnitude of AGN luminosity, which can be roughly described as $\tau \sim 200\times (L_{\rm bol}/10^{11}L_\odot)^{0.5}$ days. On the secondary y-axis on the right, we also indicate the corresponding sizes of the dust emission regions assuming $R=\tau c$, where $c$ is the speed of light.
}
\label{fig:lags}
\end{center}
\end{figure}

All of these studies are consistent with the hottest dust, i.e., the innermost region of the dusty circumnuclear torus, lying near the graphite  sublimation distance from the accretion disk. This behavior is illustrated in Figure~\ref{fig:lags}. Despite the different data and analysis used by different authors, the general trend between the near-IR dust time lag (i.e., the hot dust size) and AGN bolometric luminosity extends over 5 orders of magnitude of AGN luminosity and it follows the expected relation as $\tau\sim 200\times (L_{\rm bol}/10^{11}L_\odot)^{0.5}$ days with a relatively low dispersion ($\lesssim$0.2--0.3 dex). This conclusion is now strongly supported by interferometric data at 2 $\mu$m for many nearby AGNs that find structures --- in some cases even resolved as rings --- at the expected radius (Figure~\ref{fig:lags}, right axis) with some additional structures in a few cases \citep[e.g.,][]{gravity2020a, gravity2020b, gravity2021}.  

Various other constraints have been reported in some studies. We will evaluate them after discussing the IR reverberation behavior of NGC 4151, which has by far the most extensive set of measurements (next section). It provides a prototype that can be helpful in evaluating the measurements of other AGNs. 

\subsection{NGC 4151 as a Prototype}\label{sec:ngc4151}

\begin{figure}[htp]
    \begin{center}
  \includegraphics[width=1.0\hsize]{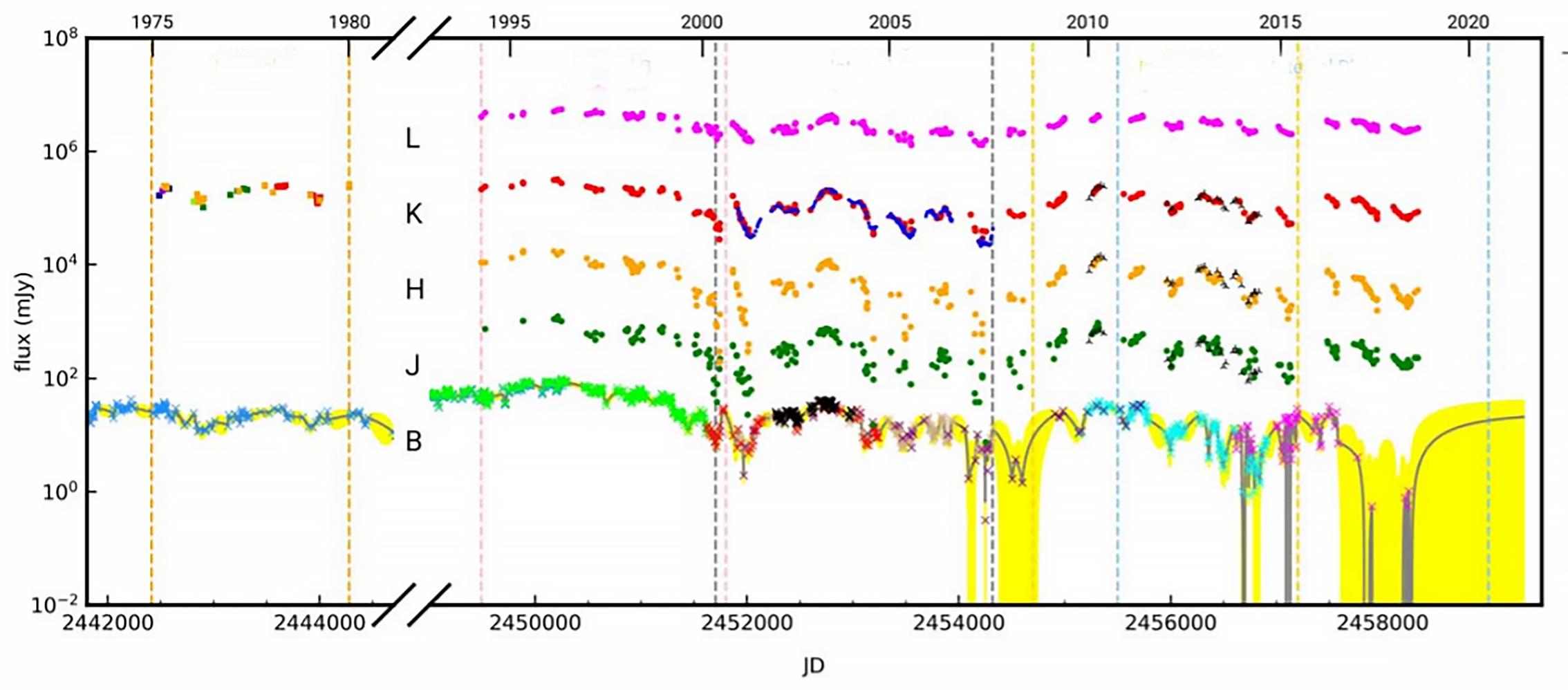}
    \caption{Historical light curves of NGC 4151 compiled from \citet{lyu2021}. For details on the data sources and the adjustment of the data to make all the sources consistent, see that paper.  
    }
  \label{fig:ngc4151}
    \end{center}
\end{figure}

NGC 4151 is unique in the extensive IR monitoring available, over 30--40 years,  see Figure~\ref{fig:ngc4151}. Nonetheless, the full potential has not been realized because each group monitoring the behavior has tended to analyze its data alone without merging into a comprehensive dataset. \citet{lyu2021} integrated all the data and analyzed the AGN behavior from 1 through 34 $\mu$m. 

\begin{figure}[H]
\centering
  \includegraphics[width=.9\hsize]{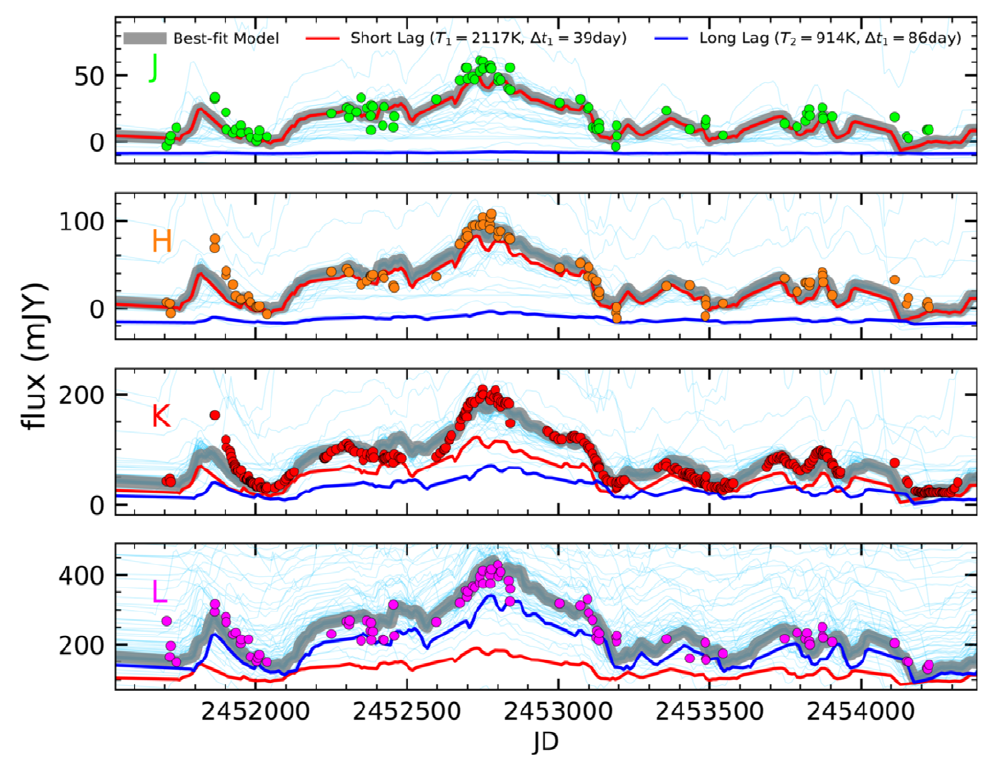}
    \caption{All the near-IR light curves for NGC 4151 in the best-sampled interval, fitted with an integrated two-black-body dust reverberation model. The observed light curves are shown as dots with different colors. The best-fit model is shown as a thick grey line and
the contribution from the two different dust components as red and blue lines. The relative contribution of the long lag (blue line) gradually
increases towards the longer wavelength and finally dominates the L-band variability. Based on \citet{lyu2021}.
    }
  \label{fig:ngc4151-2lag}
\end{figure}

For the near-IR reverberation signals of NGC 4151, \citet{lyu2021} found  two distinct dust lags of $\sim$41 and $\sim$90 days with 
the latter contribution increasing as a function of wavelength and becoming dominant in the $L$-band (see Figure~\ref{fig:ngc4151-2lag}). \citet{lyuB2022} recently measured the WISE 3.4 and 4.5 $\mu$m dust time lags of NGC 4151
to be $\sim$75--100 days, consistent with the longer lag. Previous near-IR reverberation mapping 
of this object had led to the conclusion that the lags are wavelength-independent from $J$ through the $L$ 
band \citep{oknyansky2015}. The contradiction is not complete; that work found lags generally clustered around 
the same two values, with 4/7 of those longer than 80 days associated with $L$ and only  2/9 cases with lags 
$<$80 days in this band. In addition, previous works report substantial changes in the time delay as summarized 
for NGC 4151 \citep{koshida2009} and it and other galaxies in \citet{oknyansky2015}. In contrast, \citet{lyu2021} 
found relatively little variation over the same time interval and using the same data (see their Table 6). Again, 
the contradiction is not complete --- all studies tend toward lags close to 40 days. To probe the cause of the 
discrepancies, \citet{lyu2021} took the period when the galaxy was being monitored by {\it both} the SAI and 
MAGNUM groups (generally providing Nyquist or sometimes better sampling) and repeated their analysis using 
only the SAI data, dropping the sampling cadence to sub-Nyquist. There was a significant effect on the 
lags, changing from 41 and 90 days to 30 and 48 days. As expected, adequate sampling is critical to obtaining 
reliable results.  

The detailed  monitoring of NGC 4151 shows the 
potential of multi-wavelength IR reverberation mapping to obtain physical insights about AGNs. 
\citet{lyu2021} identify five source components: (1) one with a lag $\sim$ 41 days and a
temperature $\sim$ 1500 K, probably associated with graphite dust at its sublimation temperature and  
presumably defining the inner rim of the dusty circumnuclear torus at $\sim$ 0.033 pc from the central engine; (2) another 
with a lag of $\sim$ 90 days and temperature of $\sim$ 900 K, probably dominated by silicate dust 
at its sublimation temperature (see Section~\ref{sec:twolags}) at a distance of $\sim$ 0.076 pc; 
(3) a persistent component at $\sim$ 700K that changes output only over long timescales --- over 
most of the duration of the study the emission from this component increased by $\sim$ 4\%/year, 
but prior to that time it underwent a decrease from the 1980s to the early 1990s \citep{oknyansky1999}: 
it is probably emission by dust buried in the torus; (4) a component with a lag of $\sim$ 3000 days 
corresponding to a distance from the central engine of $\sim$ 2.2--3.3 pc and at a temperature 
of $\sim$ 285 K, possibly associated either with the outer part of a flared circumnuclear torus 
or as part of a polar wind; and (5) a non-variable component that dominates the output at $\lambda\ge$ 20 $\mu$m. 

These conclusions correlate with structures deduced from the SED, imaging, and interferometry. 
The placement of the inner edge of the dusty torus at the graphite sublimation radius agrees with the 
conclusions of  \citet{barvainis1987, mor2012} from modeling of the SED, and with interferometric  measurements  \citep{honig2014}. The long reverberation lag at 10 $\mu$m agrees with the size 
of the source at this wavelength measured with interferometry \citep{burtscher2009}. This behavior 
clarifies the relation of these structures to each other and their overall placement within AGNs. The large difference in timescales between the lags around 2 $\mu$m (41 days) 
and the lag at 10 $\mu$m ({$\sim$ 3000 days}), plus the amplitudes of the lagging source components, 
presents a challenge to purely clumpy models for the torus, which generally predict only modestly 
longer timescales and lower amplitudes for 10 $\mu$m variations compared with those at 2 $\mu$m 
\citep{almeyda2017,almeyda2020}. This discrepancy is discussed in Section~\ref{sec:clumpchallenges}.  In comparison, the time scales are 
readily consistent with the wind models proposed by \citet{honig2013}, although the spectrum of 
the AGN does not show the expected strong silicate emission. Before discussing these issues, we extend the discussion to AGN variability in general.

\subsection{Other Observation of AGN IR Variability}

\subsubsection{General Lack of Classical AGN  Variability at 10--24 $\mu$m}\label{sec:var-20um}

If the mid-IR flux of classical AGNs arises through dust in a wind or outer zones of a torus reprocessing the nuclear UV-optical luminosity, there is a prediction that the components dominating the output at 
the longer mid-IR wavelengths should have very long reverberation-type variability timescales, 
as demonstrated for NGC 4151. For sources more luminous than that example, the outputs 
at 10--20 $\mu$m should appear to be non-variable over any reasonable measurement period.  
This prediction has been confirmed by \citet{lyu2019}. They took advantage of the ability 
of the MIPS instrument on {\it Spitzer} to provide photometry repeatable to the $<$ 1\% level 
in observations at 24 $\mu$m. They found that, except for blazars and flat-spectrum radio 
sources whose mid-IR emission could be non-thermal, the majority of
AGNs have typical variation amplitudes at 24 $\mu$m of no more than 10\% of that in the WISE W1 band, indicating that the
dust reverberation signals damp out quickly at longer wavelengths. In particular, steep-spectrum radio quasars also
lack strong 24 $\mu$m variability, consistent with the unification picture of radio-loud AGNs. 

\subsubsection{Ubiquitous IR Variability of Steep Spectrum Radio Sources}

Although they are not within the scope of this review, we remark that flat spectrum radio sources - a category that includes blazars and their relations - are very commonly, perhaps ubiquitously, variable across the mid-IR \citep{lyu2019}. This behavior is consistent with the unification of radio-loud AGNs \citep{urry1995} that the jet emission of these objects has been greatly amplified due to beaming effects with a very small inclination angle.

\subsubsection{Near-IR SED of the Variable Nonthermal AGN Continuum}

In the near IR, accurate determination of the emission by the dust depends on being able to remove the nonthermal continuum associated with the central engine. Theoretically, the slope into the IR is expected to be $\propto \nu^{1/3}$ \citep{shakura1973}. However, it has been difficult to confirm this behavior in the optical \citep{gaskell2008}, and thus, there is no obvious reason to expect it in
the IR. \citet{kishimoto2008} tested for this behavior on six luminous quasars, finding a spectrum $\propto \nu^{0.41 \pm 0.11}$. \citet{garcia-bernete2019} used very small aperture photometry and found slopes consistent with the theoretical value.  \citet{lyu2021} analyzed the \citet{schnulle2015} data that gave simultaneous coverage of the nonthermal regime at the z band
(0.9 $\mu$m) and the JHK bands extending into the thermal one and found evidence for a steeper-than-theoretical slope for the nonthermal spectrum. Fortunately, the differences among these possibilities are modest in the 1--1.3 $\mu$m range, and at longer wavelengths the radiation by the dust overwhelms the contribution by the nonthermal continuum in any case.  

\subsubsection{Presence of Two Lags in Near-IR Reverberation Signals of AGNs} \label{sec:twolags}

The influence of graphite at its sublimation temperature in the torus SED is well established. The second lag observed for NGC 4151 might be associated with a range of possible torus structures, although its relative lag suggests an association with sublimation of silicate grains. A test of this possibility is whether there is a similar second lag in other AGNs, differing in lag time by  about the same ratio as the two in NGC 4151. From experience fitting NGC 4151, at K-band the longer lag component should still account for a significant fraction of the total signal. The photometry of \citet{koshida2014} and \citet{minezaki2019} is sufficiently accurate to allow a search for such a double lag. It is necessary to discard cases with inadequate data, light curves that clearly depart from the reverberation mapping paradigm (e.g, large departures between the delayed B-band and K-band curves), and cases with poor fits. Nonetheless, the fits are successful for 14/17 of the \citet{koshida2014} sample and 19/30 of the \citet{minezaki2019} one, with overall relations to the trend of $\Delta \tau$ with bolometric luminosity of

\begin{equation}\Delta \tau_\textrm{$K$, long} = 10^{2.14\pm0.02} \left(L_{\rm bol}/10^{11} L_\odot \right)^{-0.5} 
\end{equation}
\noindent
and

\begin{equation}
\Delta \tau_\textrm{$K$, short} = 10^{1.77\pm0.02} \left(L_{\rm bol}/10^{11} L_\odot \right)^{-0.5} 
\end{equation}
\noindent
The ratio of short to long lags is then $0.43 \pm 0.03$, identical to that for NGC 4151, which is 0.41. The longer lags are also consistent with the reverberation results at $\sim$3.4 and 4.6~$\mu$m \citep{lyu2019}, where the corresponding dust component becomes more dominant. It appears that the two-lag behavior, with a similar difference in the lags, is a general phenomenon, justifying the connection with silicate grain sublimation. That is, the inner zones of the circumnuclear tori in many/most/all(?) AGNs have structures likely governed by the sublimation of graphite and silicate grains.

\section{Completing the AGN Census with IR Selection}\label{sec:census}

Optical spectroscopic surveys such as SDSS enable the identifications of large numbers of AGNs through their unique optical/UV emission line characteristics. They have been supplemented by  selections using distinctive 
characteristics of AGN SEDs: UV excess; X-ray emission, particularly with a hard spectrum; or variability in the optical/UV. 
All these approaches are, unfortunately, biased against heavily obscured samples \citep{hickox2018}. 
The unique IR SED features of AGNs, as compared with quiescent and star-forming galaxies, are central 
to completing the census of the whole AGN population through discovery of representative obscured objects.

\subsection{Selection via Color-Color Diagrams}

It was apparent early-on that AGNs have  characteristic IR behavior. However, a sensitive all-sky survey covering the mid-IR was required to identify AGNs using this behavior, which was provided for the first time by IRAS. These observations were used to detect luminous AGNs through the unique ``warmer'' SEDs they exhibited compared with star forming galaxies \citep[e.g.,][]{degrijp1985, osterbrock1985}. The 2MASS survey provided another opportunity to identify AGNs through photometric colors;  \citet{cutri2000} reported the discovery of a large number of AGNs from 2MASS data. These methods were extended further into the IR with ISO data \citep{laurent2000,leipski2005}. 

{\it Spitzer} launched a variety of  methods to use IR photometry to identify AGNs, most notably IRAC color-color diagrams proposed by \citet{lacy2004} and   \citet{stern2005} or, at low redshift, just the 3--5 $\mu$m color \citep[e.g.,][]{ikiz2020}.  \citet{veilleux2009} proposed a color-color selection based on longer wavelength continua measured spectroscopically. An alternative approach used fitting to the IRAC colors to identify objects with power law SEDs \citep{alonso2006, donley2007}.  The color and power law methods were merged by \citet{donley2012}, who proposed a more constrained selection criterion in the form of a power law zone within the previously proposed color-color diagrams. These methods have proven quite powerful; however, because the longest IRAC band is centered at 8 $\mu$m, they lose sensitivity to obscured AGN at modest redshift, and  they are best suited to find cases with typical Type-1 SEDs or modestly obscured Type-1 behavior \citep{donley2012}. To improve on this situation,  \citet{alonso2006, donley2007} selected in IRAC colors and then checked 24 $\mu$m and argued that there was an obscured population from the full SED. Using the R$-$[4.5] color, \citet{hickox2007} found a significant population of obscured AGN in the Bootes field. At the same time, they show (their Figure 9) how there is a strong bias against strongly obscured AGN even at moderate redshift. Surprisingly, even with this limitation a large number of sources are found that are very faint in the X-ray, indicative of strong absorption there despite the lack of evidence for strong obscuration in the optical and near IR \citep[e.g.,][]{donley2012, delmoro2016}. 

The WISE W3 band at 12 $\mu$m improved the wavelength baseline for finding AGNs \citep{assef2018}, with less bias against obscured nuclei \citep[e.g.,][]{jarrett2011,stern2012, mateos2012, hainline2014, secrest2015, hviding2018, asmus2020, carroll2021}. Recent improvements in this approach are described by \citet{hviding2022}, based on an extensive study of AGNs identified through optical spectroscopy. This approach allows them to optimize the color selection method so it can be applied in cases too obscured for reliable optical identification and to push the color selection down to lower AGN luminosities.  Combining hard X-ray all-sky surveys with WISE or other mid-IR data has resulted in significant advances in finding heavily obscured AGNs locally \citep[e.g.,][]{gandhi2009,mateos2012,asmus2020}, although it is believed that the sample is still incomplete \citep[e.g.,][]{asmus2020,carroll2021}. WISE mid-IR measurements have also proven useful to locate and confirm Compton thick AGNs in the NuSTAR data  \citep[e.g.,][]{boorman2016,lamassa2019}. The addition of the two mid-IR bands from Akari provided more possibilities in this direction \citep{lam2019}. Reliable selection can be achieved by combining mid-IR photometry from IRAC and MIPS on {\it Spitzer} with far-IR photometry from PACS and SPIRE on {\it Herschel} \citep{kirkpatrick2013}.   However, all of these methods rely on datasets that are relatively shallow, limiting the results to either very luminous or relatively nearby examples. 

The color-color and power law fitting techniques have the major advantage that they are simple and involve a minimum of free parameters. Given that they are being used with a limited set of mid-IR photometry, these features are important for returning unambiguous results. Nonetheless, there is a range of interpretations of the number and characteristics of AGNs found in this manner, as discussed in the next section.  Proposals have been made to use similar approaches with the more extensive photometry that will be provided by the James Webb Space Telescope (JWST) \citep[e.g.,][]{messias2012, kirkpatrick2017}.

\subsubsection{Limitations of AGN IR color selection}

IR color selection is a powerful technique to identify AGNs, including many not revealed by other methods. Used conservatively it can be quite accurate \citep{satyapal2018}. Of course, the approach also has limitations. The methods are not very sensitive to low luminosity AGNs that might not stand out from the host galaxy colors. In addition, with deep IRAC data the color selection becomes degenerate with stellar continua at high redshifts \citep[e.g.,][]{donley2008, mendez2013}. The limited ability to characterize SEDs is also a weakness. Finally, most popular color selection methods are unable to identify heavily obscured AGNs, but are focused on power law SEDs with perhaps only modest reddening.  Methods extending into the far-IR avoid some of these issues \citep{kirkpatrick2013} but are limited by the availability of observations of sufficient depth. 

The interest in IR selection largely arises because of the possibility of finding additional highly obscured cases. So far the results are inconclusive. For example, \citet{mendez2013} concluded that only 10\% of IR-selected AGNs are heavily obscured, but this result appears to be contradicted by \citet{delmoro2016}, who concluded that 30\% of the IRAC-selected AGN were too obscured in the X-ray to be detected there. Rather than viewing the difference as a contradiction, it is a reflection of our ignorance and it is better to take both numbers as probable lower limits. Indeed, the predictions of a very large population of heavily obscured AGN \citep[e.g.,][]{ananna2019} suggest that the current methods are only finding a small fraction. The more extensive JWST color information may mitigate these issues, but using simple color-color diagrams will be a relatively primitive approach to the extensive dataset that will become available. 

\subsection{Selection by SED analysis} \label{sec:SEDanalysis}

In principle, a more powerful approach is to fit the photometry with SEDs representing the various underlying components --- stars, AGNs, star-formation powered IR excesses, extinction. Models of this type have been demonstrated by \citet{fan2016, brown2019, azadi2020,poul2020}. However,  as discussed in Section~\ref{sec:sedfits}, these models are characterized by large numbers of free parameters, e.g., a large number for the star forming SEDs and 10 AGN SEDs for \citet{brown2019}, or 23 free parameters listed by \citet{poul2020}, and/or they are strongly based on theoretical models that themselves are subject to many free parameters and are not constrained by prior  matching to observations of well-studied AGNs \citep{brown2019,azadi2020}.  These approaches are best suited to characterizing AGNs measured at good signal to noise in multiple independent photometric bands to help remove degeneracies in the fits.

\subsubsection{Inputs to SED Fitting} \label{sec:SEDinputs}

In general, the data to be fitted with model SEDs to search for new AGNs
consist even in the most favorable case of a modest set of photometry, with
many measurements linked by the underlying physics and not truly independent.
Therefore, {\it the ability of such fitting approaches to provide unambiguous
answers depends critically on selecting inputs that provide accurate fits with
a minimum of free parameters.} Given the importance of this statement, we
describe the approaches in detail. 

\citet{yang2021} address an approach to a model for JWST data but do not apply
it to existing data. They used X-CIGALE to provide a library of non-AGN galaxy
SEDs. They then added an AGN using theoretical predictions from
\citet{stalevski2012, stalevski2016}, and subjected the resulting synthetic
images to JWST/MIRI simulated photometry. They used X-CIGALE for Markov Chain
Monte Carlo SED fitting to the results to extract the AGNs. Their approach is
thus similar to the other SED-fitting efforts in the number of free parameters
and hence risk of degeneracies, as discussed in Section~\ref{sedfitting}. 

\citet{lyu2022a} took a much more aggressive approach to limiting the free
parameters in their fits.  They used existing capabilities in the Bayesian
fitting program Prospector \citep{leja2017} to fit the stellar photospheric
continua. In the GOODS-S region where they have applied this approach, the data
over the relevant spectral range are extensive and Prospector is thoroughly
optimized to provide accurate fits. Importantly for identifying obscured AGNs
in the mid-IR, the stellar photospheric SEDs of galaxies are very uniform from
1--5 $\mu$m regardless of the galaxy type \citep[e.g.,][]{willner1984,
mannucci2001, brown2014}.  

In the mid-IR spectral region of interest for the models, the dominant feature
in star-formation-heated dust is PAH bands, and in fact the bands short of
$\sim$ 15 $\mu$m, since the longer-wavelength SED is shifted beyond reach of
the {\it Spitzer} 24~$\mu$m band (and of JWST/MIRI) at the redshifts of
interest.  At the redshifts ($\sim$ 2)  and luminosities of interest, the
variety of star-formation-heated far-IR SEDs is much less than locally
\citep{rujo2011, rujo2013, shipley2016}.  Therefore, \citet{lyu2022a} have
taken the star-heated dust  spectrum from the log(L(IR)) = 11.25 template of
\citet{rieke2009}, which provides a good fit to the overall IR SED of
high-luminosity star forming galaxies at the relevant redshift
\citep{derossi2018}.  For $\sim$ solar metallicities,  the 6--12 $\mu$m PAH
bands are usually similar in relative strength; typical variations in relative
band strengths are $\sim$30\%, 1-$\sigma$ \citep{smith2007}
.
The general behavior of these bands has little dependence upon metallicity
(although the 17 $\mu$m complex does change significantly), as can be seen
comparing the spectra presented in \citet{hunt2010} with templates determined
from solar metallicity galaxies  \citep{rieke2009}. In addition, for the narrow
range of luminosities of interest and in the 2--15 $\mu$m range, the continua
underlying the PAH bands are very similar, independent of luminosity
\citep{dacunha2008,rieke2009}.  Therefore,  it is possible to fit the
star-formation-heated dust emission with a single free parameter, i.e., the
normalization of the spectrum. 

For AGNs, the  2--5 $\mu$m region is critical, where the flux from star
formation is low but there is significant flux from AGNs
(Figure~\ref{fig:sed-decomp2}).    \citet{lyu2022a} use the ``normal'' quasar
SED to minimize free parameters and because it is an appropriate average
(Section~\ref{sec:polardust}).  The application of the AGN-based attenuation
curve (Section~\ref{sec:obscured})  allows them to fit a full range of
obscuration with a single additional free parameter.  In summary, their AGN
fits depend on two free parameters --- the normalization of the intrinsic SED
type and the amount of obscuration.\footnote{With the additional constraints
with the multiband JWST MIRI measurements, more flexibility can be added to the
fits, for example to include a HDD template and a term for polar dust.}

\subsubsection{Results}

\begin{figure}[htp]
    \begin{center}
  \includegraphics[width=1.0\hsize]{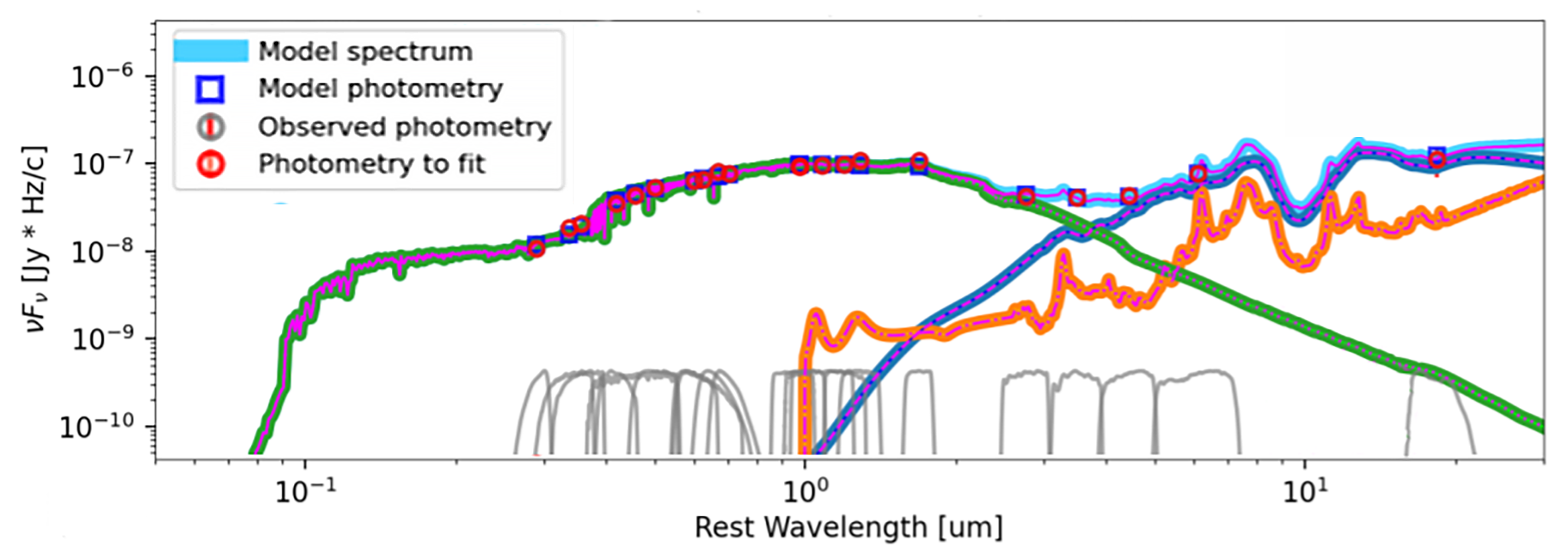}
    \caption{Heavily Obscured AGN Fit for 3D-HST 10260 (z = 0.279), from
	    \citet{lyu2022a}. For this galaxy, the stellar photospheric
	    emission is dominant at wavelengths short of 2 $\mu$m. The
	    measurements at 4.5--24 $\mu$m cannot be fitted by a combination of
	    stellar and PAH band  emission, as would be expected for a
	    star-forming galaxy, because a normalization of the star-forming
	    template in the 4.5--8 $\mu$m range makes it exceed the 24 $\mu$m
	    point by nearly a factor of three. The strongly obscured AGN
	    template (heavy blue line) is a far better fit; it drops
	    sufficiently rapidly toward shorter wavelengths to be insignificant
	    in the optical and ultraviolet.}
  \label{fig:cartoon}
    \end{center}
\end{figure}

Figure~\ref{fig:cartoon} illustrates a fit from \citet{lyu2022a}. The elevated
points above the photospheric level in the 3--6 $\mu$m range are an unambiguous
signature of an embedded AGN; the star-forming PAH-dominated template
normalized to account for these points exceeds the 24 $\mu$m point
substantially. \citet{lyu2022a} found their SED analysis was responsible for
$>$ 40\% of the identified AGN candidates, many of them unique to this method. 

The source in Figure~\ref{fig:cartoon} is a counterexample to the comment by
\citet{hickox2018}: "A surprising result from the analysis of the mid-IR
continuum of AGN is that obscured AGN have broadly similar mid-IR SEDs to
unobscured AGN." The SED fitting indicates that as many as $\sim$ 20\% of the
indicated AGNs are sufficiently strongly obscured to show strong effects in the
mid-IR SEDs. If confirmed once we have the more detailed JWST data between 5
and 25 $\mu$m, this would indicate that the lack of obscured SEDs is a
selection bias and that there are many more such AGNs to discover.  

\subsection{Selection by IR variability} \label{sec:varselect}

AGN variability across the full spectrum is a huge subject and most of it is
outside the scope of this review. However, here we discuss use of near- to
mid-IR variability to supplement optical/UV variability surveys
\citep{ulrich1997}, providing another way to identify AGNs. As summarized in
Section~\ref{sec:variability}, although the longer IR wavelengths tend not to
vary on useful timescales, the AGN near-IR (1--5 $\mu$m) emission contains the
signals from the dust reverberation response to the accretion disk variability.
Given that the IR wavelengths are relatively less obscured than the optical
band, IR variability is expected to be very useful to identify obscured AGNs.

A pioneering investigation of this type is reported by \citet{secrest2020}, who
ironically found that variability is not a good indicator of AGNs in dwarf
galaxies, possibly indicating a fundamental difference in their AGNs from those
in more massive host galaxies.  Early efforts with {\it Spitzer} include IRAC
3.6 and 4.5 $\mu$m and MIPS 24 $\mu$m variability selections in some deep
survey fields where multiple exposures have been carried out
\citep[][]{kozlowski2010, garcia-gonzalez2015, kozlowski2016}.
\citet{kozlowski2016} found a large number of AGNs using repeated IRAC
photometry of the nine square degree Bo\"otis survey. A detailed survey of
multiple fields surveyed with IRAC is described by \citet{polimera2018}. At
typical redshifts for their targets, the rest wavelengths are generally in the
1--3 $\mu$m range, i.e., variability could arise either from the nonthermal
emission or from the reradiation by dust.  They found $\sim$ 1\% of the
galaxies in multiple fields are variable, indicative of AGNs.   

 With the time-domain surveys made accessible by the WISE/NEOWISE mission and
 groundbased JHK surveys, more authors have explored use of IR variability to
 find AGNs \citep[e.g.,][]{assef2018, ward2021}. Recently, \citet{elmer2020}
 explored high-$z$ AGN selection with  long-term near-IR variability using
 archival UKIDSS data. This is an example of a number of investigations with
 groundbased JHK data; typically at the redshifts of the sources these studies
 are investigating the nonthermal source variations rather than true rest-IR
 variability. 

\citet{sheng2017}  showed that the transitions between broad H recombination
lines and weak ones in changing look AGNs are accompanied by large changes in
the fluxes in the WISE bands (see also \citet{assef2018}).  This behavior
indicates that the IR fluxes are reradiation by circumnuclear dust of energy
absorbed from variations in the accretion disks, as confirmed by their
reverberation lags \citep{lyuB2022}. The alternative explanation for the
hydrogen line changes, changing  obscuration, is not likely given that even
larger changes would be expected in the optical continua and furthermore that
the orbital period for a cloud to occult the circumnuclear torus is much too
long \citep{lamassa2015,sheng2017}.  \citet{stern2018} discuss causes for the
changing look behavior further.  WISE data are being used to find these
``changing look'' AGNs \citep{stern2018,sheng2020}. 

JWST and the Nancy Grace Roman Space Telescope will enable long-term
variability determination through comparison of fluxes with these missions to
those measured with {\it Spitzer} and groundbased near IR surveys,
respectively. There are some lessons learned for these future studies from
optical variability monitoring.  An alarming result is that methods of
increased sophistication have failed to confirm a large majority of apparently
variable AGNs, even using virtually the same optical dataset
\citep{pouliasis2020}. This may result from the effects of low-level electron
production from cosmic ray hits \citep{hagan2021}. In any case, it indicates
the importance of self-confirmation in variability studies, e.g., by detecting
the effect in more than one spectral band and separated in time.

\section{Synthesis of Constraints on the AGN Dusty Environment}
\label{sec:synthesis}

This section integrates what we have learned from AGN SEDs and reverberation
analysis to provide insights to the dusty structures around AGN central
engines.  In the literature, quite a number of theoretical models  relate
hypothesized AGN structures and their emission in detailed ways (for impressive
examples, see, e.g., the radiative transfer study of NGC 1068 by
\citet{viaene2020}, the library of clumpy disk models by \citet{nikutta2021a},
the combined clumpy disk and wind models of CAT3D-WIND \citep{honig2017}, and
the hydrodynamic models of \citet{williamson2020}, among many others). In these
works, with the assumed dust structures and grain properties, the resulting IR
SEDs and images can be directly computed through radiation-transfer
calculations. To make an analogy, this approach is similar to imagining the
behavior of a dragon and predicting its tracks as a test.  However, such models
are typically degenerate and have been challenged by recent observations (see
discussions below). Our review will take an inverse approach by making largely
qualitative (and more limited but perhaps more certain) descriptions of the AGN
dusty structures (the dragon) based on the relevant empirical observations (the
tracks).

\subsection{The Uniformity of the 1--5 $\mu$m SEDs and the Radial Structure of the Inner Torus}

The general similarity of AGN dust emission SEDs at 1--5~$\mu$m (Section~\ref{sec:agn-sed-general-feature}) indicates similar properties 
of the inner zones of the circumnuclear torus. The same conclusion holds to $z\gtrsim6.5$, where the quasar hot dust emission SED shows no evidence for  evolution compared with low redshift AGNs. Reverberation mapping suggests that the hot inner dusty torus has two separated dust
components. 
From a theoretical viewpoint, the inner radius of the dust structures 
defined by the sublimation of graphite grains is \citep[][see also \citep{lyu2021}]{barvainis1987}:
\begin{equation}
    R_\textrm{sub, C} = 1.3 \left(\frac{L_{\rm UV}}{10^{46} {\rm erg~s}^{-1}} \right)^{0.5} \left( \frac{T_{\rm sub, C}}{1500 K} \right)^{-2.8} \left( \frac{a_{\rm C}}{0.05~\mu m}\right)^{-0.5}~~ \textrm{pc},
\end{equation}
\noindent
Similarly, for the silicate grains, we have
\begin{equation}
   R_{\rm sub, S} = 2.7 \left(\frac{L_{\rm UV}}{10^{46} {\rm erg~s}^{-1}} \right)^{0.5} \left( \frac{T_{\rm sub, S}}{1000 K} \right)^{-2.8} \left( \frac{a_{\rm C}}{0.05~\mu m}\right)^{-0.5}~~ \textrm{pc},.
\end{equation}
\noindent
If the innermost boundary is set by the sublimation of graphite dust, then we expect a second component at just over twice the radius due to sublimation of silicate grains. This prediction is consistent with the observed difference in reverberation lags of just over a factor of two (Sections~\ref{sec:ngc4151} and \ref{sec:twolags}).

Since the metallicities of material surrounding AGNs are high out to the
highest redshifts probed \citep{maiolino2019}, the relative abundance of
graphite and silicates will be similar in AGN environments. The result is that
the inner zones of the circumnuclear tori will be governed by the stability of
the grains and will have very similar structures.  That is,  the similar near
IR properties among different AGNs indicate the similar temperatures, positions
(adjusted for central engine luminosity), and compositions of the emitting
grains in the AGN sublimation zones.  However, our probing is limited because
most reverberation mapping campaigns extend only to 2.2 $\mu$m; learning more
about the inner dusty torus would benefit by extending reverberation mapping
toward 5 $\mu$m. 

\subsection{Pure Clumpy Torus Models vs. Observations} \label{sec:clumpchallenges}

The torus is expected to be clumpy, based on very general considerations of which the most compelling is the variability in absorption in the X-ray \citep{markowitz2014, laha2020}, modeled in terms of clumpy models by \citet{buchner2019}. Clumps  are also needed to account for the low dust temperatures seen close to the nuclei of AGNs \citep[e.g.,][]{poncelet2006}. This possibility was anticipated in the prescient work of \citet{krolik1988}. A closeup version, without the intense energy input from an active AGN, is provided by the clumpy, tidally sheared circumnuclear disk in the Milky Way's Galactic Center \citep{tsuboi2018}.

Purely clumpy models, i.e., little or nothing between the clumps, have been successful in fitting most features of the AGN torus SED \citep[e.g.,][]{ramos2009,ramos2011,alonso2011,lopez2018}. However, as summarized by \citet{netzer2015}: "torus models contain enough free parameters to fit almost any observed SED." Nonetheless, they have three observational shortfalls.  First, they do a poor job of fitting the 1--4 $\mu$m SED  \citep{mor2009, hernan2016}. \citet{mor2009} introduced a hot black body component associated with sublimating graphite dust to improve their fits, illustrated by the modeling of \citet{swain2021}. \citet{garcia2017} show how adjustments in the standard clumpy model can produce improved fits. \citet{almeyda2020} found that the predicted
torus sizes at 2.2 $\mu$m from traditional clumpy models are typically a factor of $\sim$ 2 larger than the constrains from
K-band reverberation mapping, consistent with the existence of a graphite-dominated torus innermost region. These results suggest that this issue can be addressed by augmenting the models, as discussed in \citet{nikutta2021a}. 

A second, and more fundamental issue, is the lack of variability at wavelengths near 10 $\mu$m and longer (Sections~\ref{sec:ngc4151} and \ref{sec:var-20um}). 
Variability at 10 $\mu$m is an intrinsic characteristic of most pure clumpy torus models. For an optically thick clump, the direct heating from the AGN is deposited within the first few optical depths, raising a thin layer on the central-engine side of the clump to a high temperature. The remainder of the clump is warmed by diffuse radiation, e.g. from this hot layer, and can have a modest temperature gradient. Thus the surface of the clump not heated directly emits at much lower temperature and can contribute substantially to the output of the torus in the mid-IR. This behavior is modeled in detail by \citet{almeyda2017} and the predicted variations at 10 $\mu$m have amplitudes a significant fraction of the amplitudes in the near IR, with modest smoothing in time. This prediction is strongly contradicted by the behavior of NGC 4151 \citep{lyu2021} and implicitly by the general lack of variations at observed 24 $\mu$m (rest $\lambda \sim$ 16--24 $\mu$m) for a large number of luminous traditional AGNs \citep{lyu2017a}. \citet{honig2011} present another study of the IR variability of clumpy torus models, based on a less detailed analytic model than the model of \citet{almeyda2017}. They predict variability at 8 $\mu$m to be an order of magnitude less than at 2.2 $\mu$m even in the favorable case of a narrow torus (i.e., density falling rapidly with radius). The behavior of NGC 4151 is also inconsistent with this prediction \citep{lyu2021}.

The third issue is that clumpy models --- indeed torus models in general ---  do not predict the correlation between the strength of the narrow line emission and the mid-IR excess (Section~\ref{sec:polardust}).  This correlation implies that a substantial fraction of the mid-IR emission originates in the NLR, not in the torus. At the same time, the persistence of significant mid-IR emission (perhaps as WDD or HDD templates) even for AGNs with virtually no narrow line emission  suggests that there may be a significant component due to the disk itself, with an augmentation by polar dust in cases with strong narrow line emission. The relative roles of the torus and the wind in the mid-IR emission is a central issue for future research. We need to develop means to separate these contributions in large, representative samples of AGNs. High resolution imaging with giant telescopes can play a central role in this goal \citep{nikutta2021b}.

\subsection{The Structure of the Torus: Clumpy, Smooth, or Both?}

\citet{krolik1988} identified the critical issue for maintaining static clumps: resisting tidal shearing and stripping in the vicinity of the black hole. If a clump is to be long-lived, it needs to be stabilized against these effects either by gravity or magnetic fields. To put some numbers on this requirement, we treat the case of gravity, following the discussion in \citet{elitzur2008}. To avoid severe shearing, the Jeans timescale for gravitational collapse should be less than the Keplerian period of the orbit. This leads to a density requirement on the cloud:

\begin{equation}
    \frac{n}{10^7 cm^{-3}} > \frac{M_{SMBH}}{1 \times 10^7 M_\odot r_{pc}^3}    
\end{equation}

\noindent
where $n$ is the density of hydrogen atoms and $r_{pc}$ is the radius of the orbit of the cloud around the black hole. Taking values appropriate for NGC 4151 just outside the silicate sublimation radius, say $r_{pc}$ = 0.1 and $M_{SMBH}$ = 10$^7$ M$_\odot$, the lower limit on the clump density is $10^{10}$ cm$^{-3}$.  

Thus stable clumps must be extremely dense and compact. Athena sprang from the skull of Zeus fully formed, but the analogy is not valid for interstellar clouds and clumps. Instead, on very basic grounds, the clumps must be extreme products of turbulence in the torus gas, with weaker clumps and a continuous gas background distributed between. 
Hybrid clumpy/smooth  models have been developed by \citet{stalevski2012, siebenmorgen2015}, which correspond more closely to this picture than either purely clumpy or purely smooth models. However, detailed modeling of the 10 and 20 $\mu$m spectrum of NGC 1068 \citep{victoria2022} combined with the high resolution imaging at 10 $\mu$m \citep{gamez2022} reveal complexities that none of these models can account for, requiring us to consider them to be useful approximations that remain a bit abstract. Indeed, it appears that none of the popular models are totally successful, even within the limited constraints, of accounting for the torus properties comprehensively \citep{gonzalez2019}.

The torus has been the subject of intense theoretical modeling  \citep[e.g.,][]{netzer2015}. 
The various hydrodynamical simulations
predict a violent
environment featuring both inflow and outflow. The simulations of \citet{wada2012}, for example, 
show a circumnuclear torus dominated by turbulence and transitory density enhancements (i.e., “clumps”)
embedded in a smoother gas distribution, with a density contrast of factor of a few. They
predict turbulence and winds continuously lifting significant
amounts of material off the torus, which would naturally explain long term trends seen for NGC 4151, such as the growing (and variable) hot
dust emission, as well as the maintenance of the polar dust. Other models \citep[e.g.,][]{honig2017,namekata2016, vollmer2018,venanzi2020, takasao2022} make qualitatively similar predictions. The models also provide a unification of hybrid clumpy/smooth torus models and the driving of winds. 
However, much of the modeling to date has concentrated on the compact, pc-sized circumnuclear torus. ALMA finds both dust and gas in a disk on much larger scales and the future should see more models linking this outer structure with the pc-sized one and simulating the flows that feed the accretion disk.

{
\subsection{Torus, Polar Dust and Dusty Narrow-line Regions}

As summarized in Section~\ref{sec:polardust}, the extended polar dust,
typically associated with the NLR, can dominate the mid-IR output of an AGN as
shown for NGC 1068 in Figure~\ref{fig:nlr_dust}. On $\sim$10--100 pc scales
(left panel), its mid-IR dust emission is very extended and elongated similar
to the NLR \citep{bock2000}.  Over $\sim$0.1--1 pc scales, the high resolution
image  \citep{gamez2022} shows a roughly conical wind rising from the vicinity
of the nucleus and sublimation zones, although even at the resolution of
MATISSE\footnote{MATISSE stands for Multi AperTure mid-Infrared SpectroScopic
Experiment on the Very Large Telescope Interferometer (VLTI), an instrument
that can resolve structures down to 3.5 milli-arcsecond. See more at
\url{https://www.eso.org/sci/facilities/paranal/instruments/matisse.html} } the
exact geometry and any contributions by disk emission are not discernible. As
another example, the MATISSE image of the  Circinus galaxy \citep{isbell2022}
shows the mid-IR to be dominated by polar emission (see also
\citep{stalevski2019}), with only some blobs of emission along the direction
expected for a circumnuclear disk. The lack of emission from the disk in this
case could, of course, result because it is shadowed by its inner structure and
not sufficiently heated to emit at 10 $\mu$m, not that the disk is absent.
 In comparison, ring-like 
dust structure is clearly revealed in NGC 1068 at shorter wavelengths (2.2
$\mu$m) \citep{gravity2020b}, indicating that the relative brightness of the
polar dust component is a function of wavelength. That is, for most AGNs the
torus-like dust structures dominate the near-IR emission, as is supported by
dust reverberation analysis and interferometry observations in the near-IR
(Section~\ref{sec:rm-results}), while the polar dust emission is significant at
longer wavelengths (Section~\ref{sec:polardust}).

\citet{alonso2021} study the relation between the torus and polar dust in additional AGNs. They compare high resolution (typical beam diameters of $\sim$ 40~pc)  images at $\sim$ 10~$\mu$m with ALMA images at 870 $\mu$m at about three times higher resolution. ALMA traces the extended circumnuclear disk that delineates the orientation of the classical  torus (see Figure ~\ref{fig:firstcartoon}). In six of the twelve galaxies studied, there is extended mid-IR structure on a 50--150~pc scale, i.e. similar to that of the extended NLR, that emerges perpendicular to the plane of the torus. The presence of this emission correlates with indicators of conditions that can drive outflows \citep{venanzi2020}, i.e., moderate Eddington ratios and moderate-to-high gas column densities. 

The cases with bright NLRs but without extended emission in  \citet{alonso2021} have lower Eddington ratios and/or have lower gas columns that allow clearing of material through blowout. In  Section~\ref{sec:polardust}, we also found that about 1/3 of AGNs, particularly those with weak NLR emission, also have much fainter - or absent - polar emission. Understanding what underlies the range  from those with mid-IR dominated by polar dust to those where this dust plays a much less prominent role is a major challenge to detailed radiative modeling of the AGN IR emission.

\begin{figure}[htp]
\centering
  \includegraphics[width=1.0\hsize]{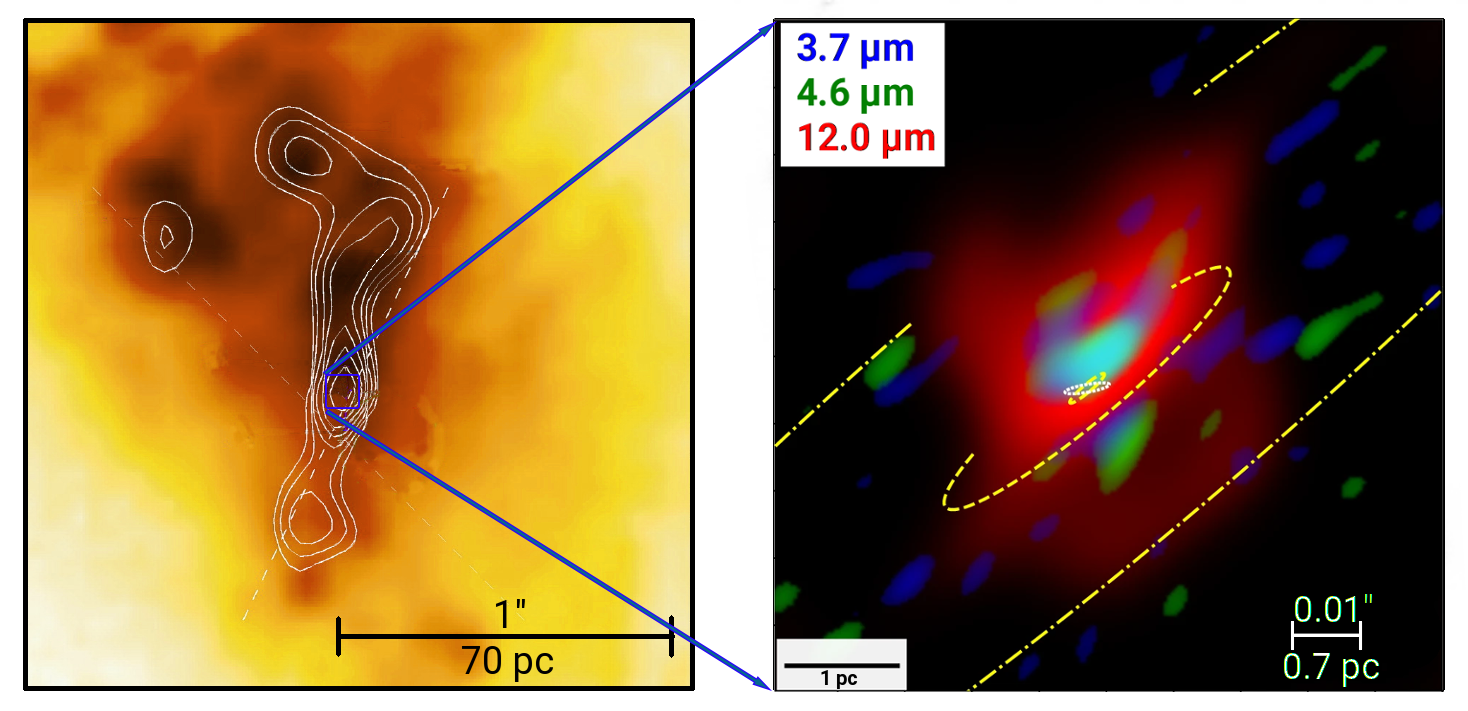}
    \caption{Optical \OIII~images overlaid with contours of the extended mid-IR
	dust emission (at $\lambda=12.5~\mu$m) of the archtypal Type-2 AGN in
	NGC~1068 (left panel) and the near- to mid-IR reconstructed images from
	MATISSE interferometry observations of the same nucleus at bands $L$
	(3--4~$\mu$m), $M$ (4.5--5~$\mu$m) and $N$ (8--13~$\mu$m) (right
	panel). Modified from the original figures published in
	\citet{bock2000} (their Fig. 4) and \citet{gamez2022} (their Fig. 1e).
	[left] © AAS, reproduced with permission; [right] reprinted by
	permission from  Nature/Springer/Palgrave).   
    }
  \label{fig:nlr_dust}
\end{figure}

\subsection{Torus Evolution and Its Relation to the Central Engine and Host Galaxy}

Dust reverberation analysis and spatially resolved observations show that the AGN torus plus disk covers physical scales bridging from the black hole accretion zone  ($\lesssim$0.01 pc) out to the host galaxy ISM ($\gtrsim$100 pc).
Structures on this latter scale are indeed seen in the submm \citep{garcia2021}, although their exact nature is likely to be complex with a variety of emission sources playing a role \citep[e.g.,][]{alonso2019,garcia2019}. 
 To feed the accretion disk and grow the SMBH, the dusty clouds within the torus have to migrate inward,  become turbulent, and be tidally disrupted. This process can naturally produce the AGN broad line emission \citep{wang2017}. 

The mass accretion rate of the SMBH can be estimated by
\begin{equation}
    \dot{M}_\textrm{BH} = 0.06\left(\frac{0.1}{\epsilon}\right)\left(\frac{L_\textrm{bol}}{10^{11} L_\odot}\right) M_\odot yr^{-1}
\end{equation}
 \noindent
 where $\epsilon$ is the efficiency of conversion of mass to energy and $L_{bol}$ is the resulting black hole luminosity. An average extended disk plus torus mass of $\sim 6 \times 10^5$ M$_\odot$ has been determined from ALMA measurements \citep{garcia2021}. A typical AGN duty cycle is $\sim 2 \times 10^6$ years \citep{khrykin2021}. Therefore, for an average luminosity for local AGNs of $\sim 10^{11}$ $L_\odot$, about 20\% of  the total disk plus torus mass eventually needs to be accreted by the black hole, a plausible value.  The issues linking the torus as discussed here and the larger gas structures feeding it and thus the black hole are reviewed in detail by \citet{honig2019} and beautifully revealed by \citet{garcia2021}.

 AGNs are proposed to be borne in ultraluminous IR galaxies and emerge as the star formation dies out \citep[e.g.,][]{sanders1988, hopkins2006}. The discovery of Compton-thick sources in the nuclei of many luminous IR galaxies \citep{ricci2021}  confirms the basic hypothesis.  
 Predictions of how AGNs and their host galaxies co-evolve have suggested feedback from AGNs, typically
associated with (dusty) winds along the AGN polar direction, is the key to regulate 
the galaxy-BH relation \citep[e.g.,][]{hopkins2012, hopkins2016, trebitsch2019}. 

However, it is a bit embarrassing that by-and-large the hosts of non-obscured AGNs appear to be normal main-sequence star forming galaxies. They have at most only a slight tendency to be disturbed or have asymmetric morphology compared with non-AGN galaxies, and their star formation appears neither to be quenched nor substantially enhanced. Efforts to deduce a time-sequence for AGN evolution have been generally unsuccessful \citep[e.g.,][]{xu2015b,hatcher2021}. How the AGN dusty and obscuring structures evolve is also an open question. Important clues may lie in an as-yet undiscovered population of heavily obscured AGNs. We should find many of these AGNs with JWST, if they exist, and can look forward to answers to some of these questions.

\authorcontributions{The authors have contributed equally to writing this article. }

\funding{This work was supported
by NASA grants NNX13AD82G and 1255094. }

\acknowledgments{We received extensive and very helpful comments on an early
draft from Almudena Alonso-Herrero, Kevin Hainline, Luis Ho, and Robert
Nikutta. We also thank Minghao Yue, Feige Wang and Jinyi Yang for their
feedback on the readability of a later draft. We benefitted from discussions
with Anna Sajina, Alexandra Pope, and Mark Lacy, and we particularly thank
Prof. Sajina for her leadership on this special issue. This research has made
use of the NASA's Astrophysics Data System Bibliographic Services and  VizieR
catalog access tool, CDS, Strasbourg, France. This publication makes use of
data products from the Wide-field IR Survey Explorer, which is a joint project
of the University of California, Los Angeles, and the Jet Propulsion
Laboratory/California Institute of Technology, funded by the National
Aeronautics and Space Administration.}

\conflictsofinterest{The authors declare no conflict of interest. }

\abbreviations{Abbreviations}{
The following abbreviations are used in this manuscript:\\

\noindent 
\begin{tabular}{@{}ll}
AGN & Active Galactic Nucleus\\
ALMA & Atacama Large Millimeter Array \\
BH  & Black Hole \\
BLR &  Broad-Line Region \\
DOG  & Dust Obscured Galaxy \\
CCD  &  Charge-Coupled Device \\
EW  & Equivalent Width \\
FWHM & Full Width at Half Maximum \\
HDD  & Hot Dust Deficient \\
HST  & Hubble Space Telescope \\
IR & InfraRed \\
IRAC  &  InfraRed Array Camera (on Spitzer)  \\
IRAS & InfraRed Astronomical Satellite \\
IRS & InfraRed Spectrograph (on Spitzer) \\
ISM  & InterStellar Medium \\
ISO  & Infrared Space Observatory \\
JWST  & James Webb Space Telescope  \\
LINER  &  Low-Ionization Nuclear Emission-line Region \\ 
MAGNUM & Multicolor Active Galactic Nuclei Monitoring \\
MATISSE  &  Multi AperTure mid-Infrared SpectroScopic Experiment (on the VLT) \\
MIPS  &  Multiband Imaging Photometer for Spitzer  \\
NEOWISE & Near-Earth Object Wide-field IR Survey Explorer\\
NLR  & Narrow-Line Region \\
PAH  & Polycyclic Aromatic Hydrocarbon \\
PACS  &  Photoconductor Array Camera and Spectrometer (on Herschel) \\
PG   & Palomar-Green \\
PSF  & Point Spread Function \\
SAI  & Sternberg Astronomical Institute \\
SDSS  &  Sloan Digital Sky Survey \\
SED  & Spectral Energy Distribution \\
SFG  & Star Forming Galaxy \\
SFR  & Star Formation Rate \\
SMBH & SuperMassive Black Hole\\
SPIRE  &  Spectral and Photometric Imaging RecEiver (on Herschel)  \\
SWIRE  & Spitzer Wide-area Infrared Extragalactic (survey) \\ 
UKIDSS  &  UKIRT Infrared Deep Sky Survey  \\
UV  & UltraViolet  \\
VLTI  & Very Large Telescope Interferometer \\
WDD  & Warm Dust Deficient \\
WISE & Wide-field Infrared Survey Explorer \\
\end{tabular}
}

\begin{adjustwidth}{-\extralength}{0cm}
\reftitle{References}

\end{adjustwidth}

\end{document}